\documentclass[12pt]{article}

\usepackage{amsmath}
\usepackage{amsfonts}
\usepackage{amssymb}
\usepackage{graphicx}
\usepackage{grffile,cite}
\usepackage{subfig}
\usepackage{hyperref}
\usepackage{xcolor}

\usepackage{float}

\begin{document}
\thispagestyle{empty}

\begin{center}

{\LARGE\bf{Analyticity Domain for Off-shell Five-point Superstring Loop Amplitudes}}
\bigskip

{\large Ritabrata Bhattacharya}\,$^{a}$, \, \, 
and \ \ {\large Ratul Mahanta}\,$^{b}$  \, \, \\

\bigskip 
\bigskip

{\small
$^a$ Chennai Mathematical Institute,
H1, SIPCOT IT Park, Siruseri, Kelambakkam 603103, India\\[3mm]
$^b$ Harish-Chandra Research Institute, HBNI, Allahabad 211019, India \\[2mm]
}

\end{center}
\begin{center}
\it{ritabratab@cmi.ac.in, ratulmahanta@hri.res.in}
\end{center}
\bigskip 
 %
%
\begin{center} 
{\bf Abstract} 
\end{center}

\begin{quotation}
\noindent 
{\small

  In an earlier work \cite{BM2021}, we showed that for arbitrary Feynman loop diagrams (with no massless internal propagators) in closed superstring field theory, a known domain can be analytically extended by simply adjoining convex combinations of points drawn from it. Such extension yielded 350 of the well-known 370 primitive tubes for 5-point functions. In this paper, we consider the remaining 20 primitive tubes. We show that different types of convex combinations cover different sections of these tubes. We use a specific algorithm to analytically obtain 129 types of convex combinations along with their region of validity. Then, we use numerical techniques to locate points from those tubes that are not covered by them. This does not rule out the possibility of the leftover points being represented as desirable convex combinations. Indeed we give ways to improve our algorithm indicating that the leftover points can be covered.
}
\end{quotation}

\newpage
\tableofcontents
\setcounter{footnote}{0}

\section{Introduction and summary}
\label{sec:intro}

A well known non-perturbative result for local quantum field theories with a mass gap is that the off-shell momentum space amputated Green's functions are analytic as functions of complex external momenta inside the primitive domain $\mathcal{D}$\cite{Bogolyubov, Steinmann1960, Ruelle1961, AB1960, Araki1961}. The proof uses causality constraints on the position space correlators and their Fourier transforms to momentum space correlators, to derive the analyticity in momenta. For an $n$-point function, the primitive domain $\mathcal{D}$ is a collection of points $\{p\equiv(p_1,\dots,p_n)\}$ subject to two conditions. Now, defining $P_I=\sum_{a\in I}p_a$ for each non-empty proper subset $I$ of $\{1,2,\dots,n\}$, these conditions can be stated as\footnote{All the Minkowski products use space-time metric with mostly plus signature, in our convention.}
\begin{equation}
\begin{aligned}
\label{eq:PD}
  &\text{C1.}\ \ p_1+\cdots +p_n=0\ ,\\
  &\text{C2.}\ \ \text{For each}\ I,\ \ \text{if}\ \text{Im }P_I\neq0,\ (\text{Im }P_I)^2\leq0\ ,\\
  &\qquad\qquad\qquad\qquad\text{else}\ -P_I^2<M_I^2\ ,
\end{aligned}
\end{equation}
where $M_I$ is the invariant threshold mass for producing any (multi-particle) state in the collision of particles carrying total momentum $P_I$. Due to C1, there are $(n-1)D$ independent complex variables for $n$ ingoing external $D$-momenta\footnote{Due to C2, the primitive domain does not contain on-shell momenta. However, the holomorphic extension of the primitive domain includes on-shell momenta, e.g. \cite{BEG1964}.}.

The holomorphic extension of the primitive domain can be used to prove various analyticity properties of the S-matrix elements (computed as on-shell connected amputated Green's functions for $n\geq3$) of QFTs \cite{BEG1965, Bros1986, Pavlov1978, MP1978, LMMPS1979, Bros1980, MPPS1982, MPPS1984}. These derivations use properties of the primitive domain (or specific subregions) in several complex variables, and the results do not depend on the functional form of the Green's functions that are analytic on the domain. In particular, properties of a subregion of the primitive domain are useful to prove crossing symmetry of amplitudes for $2\to2$ scattering \cite{BEG1965}. Also, holomorphic extension of the same subregion leads to the JLD domain in one particular combination of external momenta $P_I$, keeping other independent combinations real at forward lightcone \cite{BMS1961, JL1957, Dyson1958}.

Local QFTs are known to be the low energy effective theories of string theory \cite{BLTP}. Thus, it is natural to ask whether the off-shell $p$-space correlators of string field theory (which has been designed to reproduce the perturbative amplitudes of string theory\footnote{See \cite{LEKSV2017} for a detailed review, and \cite{PS2016} for the momentum space Feynman rules.}) are also analytic in the primitive domain. Because the position space construction is unknown\footnote{The key obstacle is that the momentum space vertices are non-local for string field theory \cite{LES2019}.}, we must directly use the perturbative expansion of off-shell $p$-space correlators in string field theory to answer this question. For this analysis, we have to consider only the contributions to off-shell correlators coming from the Feynman diagrams in string field theory with no massless internal propagators, to avoid the issues with IR divergences as in local QFTs with massless states \cite{LES2019}. We refer this part of the off-shell correlators as the infrared safe part. When all the external particles are massless, this precisely gives the vertices of the Wilsonian effective field theory of massless fields obtained by integrating out the massive fields in superstring theory \cite{LEKSV2017, Sen2017}.

Along this line, in closed superstring field theory (SFT), \cite{LES2019} considered any Feynman loop diagram $\text{F}(p)$ with $n$-amputated external legs which does not contain any massless internal propagator, and proved its analyticity on the LES domain $\mathcal{D}'$ given by the following three conditions\footnote{The proof uses the Feynman integrals for the relevant Feynman diagrams, and deformations of loop momentum integration contours without crossing singularities (which are poles of the internal propagators) of the integrand on complex loop momenta planes.}.
\begin{equation}
\label{eq:LESD}
  \text{C1},\ \text{C2}, \text{\ and C3: each}\ \text{Im }p_a\ \text{lies on a 2d Lorentzian plane}\ p^0-p^1\ \text{if}\ \text{Im }p_a\neq0\ .
\end{equation}
Due to the additional condition C3, LES domain is a proper subset of the primitive domain. However, this is the subregion that was used to prove crossing symmetry and was analytically extended to the JLD domain. Also, there are two important corollaries of the work of \cite{LES2019}, which state
\begin{equation}
\label{eq:LEScor}
\begin{aligned}
  &\text{Cor1.}\ \ \text{F}(p)\ \text{remains analytic at}\ \tilde{\Lambda}p,\ \text{if}\ p\in\mathcal{D}'\ ;\\
  &\text{Cor2.}\ \ \text{F}(p)\ \text{remains analytic in a small open neighbourhood at each}\ p\in\mathcal{D}'\ .
\end{aligned}
\end{equation}
Where, complex Lorentz transformations $\tilde{\Lambda}$ (i.e., matrices $\tilde{\Lambda}$ with complex entries such that $\tilde{\Lambda}^T\eta\tilde{\Lambda}=\eta$, $\eta$ denotes the Minkowski metric in $\mathbb{R}^D$) act on $p$ as $\tilde{\Lambda}p\equiv(\tilde{\Lambda}p_1,\dots,\tilde{\Lambda}p_n)$. Adjoining new points allowed by the two above corollaries with $\mathcal{D}'$, we denote the resultant domain by $\tilde{\mathcal{D}}'$.

In a previous paper \cite{BM2021}\footnote{This uses no further input from SFT.}, we holomorphically extended $\tilde{\mathcal{D}}'$ to a larger subset of the primitive domain $\mathcal{D}$ using Bochner's tube theorem\footnote{Bochner's theorem is a classical result on the theory of functions of several complex variables \cite{Bochner1937}.}. This extension formula holds for any $n$-point functions. Explicit applications to 2-, 3-, and 4-point functions showed that the extension equals the primitive domain. For the case of 5-point functions, the primitive domain essentially contains 370 primitive tubes $\mathcal{T}$, which are of the form
\begin{equation}
  \mathcal{T}=\mathbb{R}^{4D}+i\mathcal{C},\quad \mathcal{C}\subsetneq\mathbb{R}^{4D}\ ,
\end{equation}
where $\mathcal{C}$ is a convex set, called the base. Each of the primitive tubes has an overlap $\mathbb{R}^{4D}+i\tilde{\mathcal{C}}$ with $\tilde{\mathcal{D}}'$, where Bochner's tube theorem applies and extends it to its convex hull, i.e. $\mathbb{R}^{4D}+i\text{Ch}(\tilde{\mathcal{C}})$. Here, $\text{Ch}(\tilde{\mathcal{C}})$ is contained in $\mathcal{C}$. And, only if $\text{Ch}(\tilde{\mathcal{C}})=\mathcal{C}$, then the full primitive tube $\mathcal{T}$ is obtained by the above extension. The bases for 350 out of the 370 primitive tubes can be brought to the following common form.
\begin{equation}
  \mathcal{C}=\bigg\{\vec{Q}=(P_\alpha,P_\beta,P_\gamma,P_\delta):\ P_\alpha,P_\beta,P_\gamma,P_\delta\in V^+\bigg\}\ ,
\end{equation}
where $P_\alpha,P_\beta,P_\gamma,P_\delta$ are respectively the imaginary parts of some four independent combinations of external $D$-momenta, and $V^+$ is the open forward lightcone in $\mathbb{R}^D$. We carried out the analysis on the above 350 primitive tubes, fully obtaining them from their overlap with $\tilde{\mathcal{D}}'$ by the aforesaid extension. Given a $\vec{Q}\in\mathcal{C}$, we found one explicit way to decompose it into four points drawn from $\tilde{\mathcal{C}}$ whose convex combination represents the $\vec{Q}$. And, this way of decomposing works for all the points in $\mathcal{C}$. The key observation was that for each of the 350 bases $\mathcal{C}$, the number of independent momenta for any $\vec{Q}\in\mathcal{C}$ (which is 4 due to external momentum conservation) is equal to the number of specific combinations of external momenta whose imaginary parts are kept in $V^+$. Due to this, the above type of convex decomposition\footnote{By a convex decomposition of a point, we mean a convex combination of several points reproducing the given point.} of a generic $\vec{Q}$ into four points drawn from $\tilde{\mathcal{C}}$ was readily discovered without using an algorithm. However, the above observation does not apply to the remaining 20 primitive tubes, whose bases (will be denoted by $\mathcal{C}'$) can only be brought to the common form
\begin{equation}
\label{eq:probtubescf}
  \bigg\{\vec{Q}=(P_\alpha,P_\beta,P_\gamma,P_\delta):\ P_\alpha,\ P_\beta,\ P_\gamma,\ P_\delta,\ (P_\beta+P_\delta-P_\alpha),\ (P_\gamma+P_\delta-P_\alpha)\in V^+\bigg\}\ .
\end{equation}
Given a $\vec{Q}$ from any of these 20 primitive tubes, finding any type of convex decomposition (if one exists) into points drawn from the overlap (will be denoted by $\tilde{\mathcal{C}}'$) will require an algorithm, which remained unresolved.

In this article, we consider the remaining 20 primitive tubes for the 5-point functions. When for a $\vec{Q}\in\mathcal{C}'$, $P_\alpha-P_\delta$ is timelike or lightlike, we explicitly give a convex decomposition of $\vec{Q}$ into four points drawn from $\tilde{\mathcal{C}}'$. There are two ways of decomposing that work for points $\vec{Q}$ with $P^0_\alpha\geq P^0_\delta,P^0_\alpha<P^0_\delta$ respectively. For all points $\vec{Q}$ with spacelike $P_\alpha-P_\delta$, again a single type of convex decomposition does not work. For the space of such points $\vec{Q}$ drawn from $\mathcal{C}'$, different ways of decomposing $\vec{Q}$ work in different portions. We provide a specific algorithm, using which we analytically obtain 129 types of convex decomposition of $\vec{Q}$ with spacelike $P_\alpha-P_\delta$ into four points drawn from $\tilde{\mathcal{C}}'$. We also analytically obtain their region of validity. The general instructions are --
\begin{itemize}
  \item[1.]{We begin by going to a Lorentz frame in which the temporal component of $P_\alpha-P_\delta$ vanishes.}
  \item[2.]{Then, we choose any four momenta from the set
  \begin{equation}
  \nonumber
    \bigg\{P_r,\ P_\beta+P_\delta-P_\alpha,\ P_\gamma+P_\delta-P_\alpha,\ P_r-P_{r'},\ r\neq r',\ r,r'=\alpha,\beta,\gamma,\delta\bigg\}\ ,
  \end{equation}
  such that the inverse transformation from them to $\{P_\alpha,P_\beta,P_\gamma,P_\delta\}$ exists. This allows to write each $P_r,\ r=\alpha,\beta,\gamma,\delta$ as a linear combination of the chosen four momenta. As a result, we can decompose $(P_\alpha,P_\beta,P_\gamma,P_\delta)$ into four terms, with all the momenta in each term lying on a 2d Lorentzian plane that will be the plane of one of the chosen momenta. E.g., if the chosen set is $\{P_\alpha,P_\beta,P_\gamma,P_\delta-P_\alpha\}$, the decomposition will be given by
  \begin{equation}
  \nonumber
    (P_\alpha,P_\beta,P_\gamma,P_\delta) = (P_\alpha,0,0,P_\alpha)+(0,P_\beta,0,0)+(0,0,P_\gamma,0)+(0,0,0,P_\delta-P_\alpha)\ .
  \end{equation}
  This way, with all consistent slections of the initial set of four momenta, we generate a list of decompositions into four terms that involve only the chosen four momenta.}
  \item[3.]{Given the list of above decompositions into four terms, we scan for only those for which one can improve the temporal components of the momenta in each of the terms such that all the four terms are brought inside $\tilde{\mathcal{C}}'$ and the terms add up to $(P_\alpha,P_\beta,P_\gamma,P_\delta)$\footnote{Analytic methods of finding convex decompositions allow improvement of the temporal components only. If one wishes to vary some or all the spatial components, then one needs numerical methods for the decomposition. Example, (this point is already covered by our analytic method)
  $$
  \begin{aligned}
  & P_1=P_2=P_3=
  \begin{pmatrix}
  1 \\
  0.098 \\
  0
  \end{pmatrix},\ \
  P_4=
  \begin{pmatrix}
  1 \\
  -0.49 \\
  0.8487
  \end{pmatrix},\\
 & (P_1\ \ P_2\ \ P_3\ \ P_4) =
  \begin{pmatrix}
0.5 & 0.5 & 0.5 & 0.99 \\
0.098 & 0.098 & 0.098 & -0.49 \\
-0.16974 & -0.16974 & -0.16974 & 0.8487 
\end{pmatrix} 
+\begin{pmatrix}
0.5  &  0.5  & 0.5  & 0.01 \\
0 &   0     &   0  & 0 \\    
0.16974 &   0.16974     &  0.16974  & 0
  \end{pmatrix}\ .
  \end{aligned}
  $$}. For each type of convex decomposition obtained this way, we determine the region of its validity where above improvement exists.}
\end{itemize}
For all decompositions obtained after the second step each term must satisfy the conditions in \eqref{eq:probtubescf} with $P_{\alpha}, \dots, P_{\delta}$ redefined accordingly for each term. This will bring all the terms inside the overlap $\tilde{\mathcal{C}}'$ which is a subset of $\mathcal{C}'$. This justifies the third step. We refer a prescription to improve the decompositions obtained after the second step (for which, it exists) as a ``consistent $\epsilon$-prescription''. The first step only helps considerably in explicitly finding such a prescription for a given type of decomposition.

Using the aforementioned algorithm, given any of the 20 remaining primitive tubes, we get a list of 129 types of convex decomposition into four points drawn from $\tilde{\mathcal{C}}'$ that are valid in different parts of the space of points $\vec{Q}$ with spacelike $P_\alpha-P_\delta$ inside $\mathcal{C}'$. We give analytical formulae that describe these parts. We employ numerics, in order to address the question of whether a point exists inside $\mathcal{C}'$ that cannot be decomposed in any of the above ways which are obtained by our algorithm. And, the answer is affirmative. Here, considering different configurations for momenta $P_\alpha,P_\beta,P_\gamma,P_\delta$ in three space-time dimensions, we numerically generate distinct points inside the primitive tube. We search for points which are not covered by above types of convex decomposition. For such a point, a convex decomposition into terms drawn from $\tilde{\mathcal{C}}'$ may still be found, which may be covered by a seperate algorithm that analytically generates new types of convex decomposition, or directly applying numerical techniques. Regarding this we provide ways to improve our algorithm to find more decompositions and show indications that the leftover points can be covered, in appendix \ref{app:algo2}.

The rest of the paper is organized as follows. In section \ref{sec:review}, we review the necessary details of the primitive tubes and the general holomorphic extension formula inside them, as well as briefly discussing their applications. In section \ref{sec:remaintubes}, we focus on the remaining 20 primitive tubes for the 5-point functions, and the holomorphic extension inside them (with a concrete realization in section \ref{sec:extensioninprobt}). Convex decompositions of points with timelike or lightlike $P_\alpha-P_\delta$ are discussed in section \ref{sec:case1}. Convex decompositions of points with spacelike $P_\alpha-P_\delta$ and our algorithm to obtain them are discussed in section \ref{sec:case2} (and in appendix \ref{app:Qdecomp}). In section \ref{sec:numerics}, we detail our numerical analysis to search for points from the tubes, that are not covered by the algorithm. Finally, we end with some discussions in section \ref{sec:discussion}.

\section{Review}
\label{sec:review}

In this section, we review the necessary details of the proposed formula in \cite{BM2021} for holomorphic extension inside each of the primitive tubes of $n$-point functions. To illustrate its applications, we briefly discuss obtention of 350 out of the 370 primitive tubes of 5-point functions using such extensions.

\subsection{Extension formula for $n$-point functions}

For $n$-point functions, the primitive domain $\mathcal{D}$ (given by \eqref{eq:PD}) essentially contains a family (indexed by $\lambda$) of mutually disjoint primitive tubes $\mathcal{T}_\lambda$ \cite{AB1960, BEG1964, Lassalle1974, BL1975}. In order to state them explicitly, we consider the space $\mathbb{R}^{n-1}$ of $n$ real variables $s_1,\dots,s_n$ linked by $s_1+\cdots+s_n=0$. Defining $S_I=\sum_{a\in I}s_a$ for each $I\subsetneq\{1,2,\dots,n\}\setminus\emptyset$, the family of planes $\{S_I = 0\ \ \forall I\}$ divides $\mathbb{R}^{n-1}$ into open convex cones with common apex at the origin\footnote{Note that $S_I = 0$ and $S_{I^c} = 0$ are the same plane, where $I^c$ denotes the complement of $I$.}. Each such cone is called a cell, inside which each $S_I$ is of definite sign $\lambda(I)$. Each such cell defines an open convex primitive tube $\mathcal{T}_\lambda$ given as\footnote{With $i=\sqrt{-1}$.}
\begin{equation}
\label{eq:primTlamb}
\begin{aligned}
  \mathcal{T}_\lambda=&\ \bigg\{p\equiv(p_1,\dots,p_n):\ {\sum}_{a=1}^n p_a=0,\ \lambda(I)\text{Im }P_I\in V^+\quad \forall I\bigg\}\\
  =&\ \ \mathbb{R}^{(n-1)D}\ +\ i\mathcal{C}_\lambda\ ,\\
  \mathcal{C}_\lambda=&\ \bigg\{\text{Im }p\in\mathbb{R}^{(n-1)D}:\ \lambda(I)\text{Im }P_I\in V^+\quad \forall I\bigg\}\ ,
\end{aligned}
\end{equation}
where $V^+$ is the open forward lightcone in $\mathbb{R}^D$. $\mathcal{C}_\lambda$ is the convex conical base of $\mathcal{T}_\lambda$, to be called the primitive cone. There are 6 cells for $n=3$, thus the number of primitive tubes is 6. It is 32 for $n=4$, 370 for $n=5$, etc.

Each primitive tube has an overlap with $\tilde{\mathcal{D}}'$\footnote{$\tilde{\mathcal{D}}'$ is defined below the equation \eqref{eq:LEScor}.}. From this overlap, we consider the tube $\bigcup_{\vec{\theta}}\mathcal{T}_\lambda^{\vec{\theta}}$ (to be called the LES tube) where
\begin{equation}
\label{eq:LESTlamb}
\begin{aligned}
  \mathcal{T}_\lambda^{\vec{\theta}}=&\ \bigg\{p\equiv(p_1,\dots,p_n):\ \forall a\ \text{Im }p_a\in\ p^0-p^{\vec{\theta}}\ \text{plane};\ {\sum}_{a=1}^np_a=0,\\
  &\quad\quad\quad\quad\quad\quad\quad\quad\quad\quad\quad\quad\text{and}\ \lambda(I)\text{Im }P_I\in V^+\quad \forall I\bigg\}\\
  =&\ \ \mathbb{R}^{(n-1)D}\ +\ i\mathcal{C}_\lambda^{\vec{\theta}}\ ,\\
  \mathcal{C}_\lambda^{\vec{\theta}}=&\ \bigg\{(\text{Im }p_1,\dots,\text{Im }p_n)\ \text{on manifold}\ {\sum}_{a=1}^n\text{Im }p_a=0\ \text{such that}\\
    &\quad\quad\quad\forall a\ \text{Im }p_a\in\ p^0-p^{\vec{\theta}}\ \text{plane}\ \text{and}\ \lambda(I)\text{Im }P_I\in V^+\quad \forall I\bigg\}\ ,
\end{aligned}
\end{equation}
and $\vec{\theta}$ is a set of $D-2$ angles specifying a two dimensional Lorentzian plane $p^0-p^{\vec{\theta}}$ with $\vec{\theta}=0$ specifying the $p^0-p^1$ plane\footnote{By acting a real rotation on $\mathcal{T}_\lambda^{\vec{\theta}=0}$, one can obtain $\mathcal{T}_\lambda^{\vec{\theta}}$. For more details, see \cite{BM2021}.}. The LES tube is given by $\mathbb{R}^{(n-1)D}+i(\bigcup_{\vec{\theta}}\mathcal{C}_\lambda^{\vec{\theta}})$. We will refer $\bigcup_{\vec{\theta}}\mathcal{C}_\lambda^{\vec{\theta}}\equiv\tilde{\mathcal{C}}_\lambda$ as the LES cone. Cor1 (given in \eqref{eq:LEScor}) guarantees the analyticity of $\text{F}(p)$ in the LES tube.

The LES tube is proven to be path-connected. It can be thickened using Cor2 (given in \eqref{eq:LEScor}) to make it open. Thus, using Bochner's tube theorem \cite{Bochner1937, Rudin1971}\footnote{The version referred to as the `convex tube theorem' at the end of the third section in \cite{Rudin1971} is appropriate for our purpose.}, the LES tube can be holomorphically extended to its convex hull given by $\mathbb{R}^{(n-1)D}+i\text{Ch}(\bigcup_{\vec{\theta}}\mathcal{C}_\lambda^{\vec{\theta}})$, where $\text{Ch}(\bigcup_{\vec{\theta}}\mathcal{C}_\lambda^{\vec{\theta}})$ denotes the smallest convex set containing the LES cone $\tilde{\mathcal{C}}_\lambda$\footnote{The primitive cone $\mathcal{C}_\lambda$ is a convex set containing the LES cone $\tilde{\mathcal{C}}_\lambda$. Thus, it contains the convex hull of the LES cone.}. This extension is non-trivial when $n>2$ for which the LES cone is proven to be non-convex. See \cite{BM2021} for proofs of the statements in this paragraph.

\subsection{Applications to Five-point functions}

For 5-point functions, there are 370 primitive tubes \cite{AB1960}. For 350 of them, the corresponding primitive cones are given by
\begin{equation}
\label{eq:5ptTbases}
\begin{aligned}
  &\mathcal{C}^{+}_a=-\mathcal{C}^{-}_a&=&\bigg\{\text{Im }p:\ \text{Im }p_b,\ \text{Im }p_c,\ \text{Im }p_d,\ \text{Im }p_e\in V^+\bigg\},\\
  &\mathcal{C}^{+}_{ab}=-\mathcal{C}^{-}_{ab}&=&\bigg\{\text{Im }p:\ -\text{Im }p_b,\ \text{Im }(p_b+p_c),\ \text{Im }(p_b+p_d),\ \text{Im }(p_b+p_e)\in V^+\bigg\},\\
  &\mathcal{C}^{+}_{ab,c}=-\mathcal{C}^{-}_{ab,c}&=&\bigg\{\text{Im }p:\ \text{Im }p_c,\ -\text{Im }(p_b+p_c),\ \text{Im }(p_b+p_d),\ \text{Im }(p_b+p_e)\in V^+\bigg\},\\
  &\mathcal{C}'^{+}_{ab,c}=-\mathcal{C}'^{-}_{ab,c}&=&\bigg\{\text{Im }p:\ -\text{Im }(p_b+p_c),\ \text{Im }(p_a+p_c),\ \text{Im }(p_b+p_d),\ \text{Im }(p_b+p_e)\in V^+\bigg\},\\
  &\mathcal{C}^{+}_{a,bc}=-\mathcal{C}^{-}_{a,bc}&=&\bigg\{\text{Im }p:\ \text{Im }p_d,\ \text{Im }p_e,\ \text{Im }(p_a+p_b),\ \text{Im }(p_a+p_c)\in V^+\bigg\}\ ,
\end{aligned}
\end{equation}
where $(abcde)=$ permutation of $(12345)$ and $\text{Im }p=(\text{Im }p_1,\dots,\text{Im }p_5)$ is linked by $\text{Im }p_1+\cdots+\text{Im }p_5=0$. Due to this link, any four linear combinations of $\text{Im }p_1,\dots,\text{Im }p_5$ which are linearly independent can be chosen as our set of basis vectors, in order to assign coordinates to the points in a primitive cone from the above list. Also, only four (out of thirty) $\text{Im }P_I$ are needed to be in specific lightcones in order to define any of the primitive cones listed in \eqref{eq:5ptTbases}. Then, for a given cone, external momentum conservation fixes all other $\text{Im }P_I$ in specific lightcones.

For each of the above 350 primitive cones, $\text{Ch}(\bigcup_{\vec{\theta}}\mathcal{C}_\lambda^{\vec{\theta}})=\mathcal{C}_\lambda$ holds, as briefly demonstrated below (for details, see \cite{BM2021}). With an appropriate choice for a basis, the primitive cones in \eqref{eq:5ptTbases} can be brought to the following common form
\begin{equation}
  \mathcal{C}=\bigg\{\vec{Q}=(P_\alpha,P_\beta,P_\gamma,P_\delta):\ P_\alpha,P_\beta,P_\gamma,P_\delta\in V^+\bigg\}\ .
\end{equation}
Now, the following simple decomposition for any $\vec{Q}\in\mathcal{C}$:
\begin{equation}
  (P_\alpha,P_\beta,P_\gamma,P_\delta) = (P_\alpha,0,0,0)+(0,P_\beta,0,0)+(0,0,P_\gamma,0)+(0,0,0,P_\delta)
\end{equation}
can be improved so that all the four terms in this decomposition are brought inside the corresponding LES cone, as demonstrated below.
\begin{equation}
\label{eq:pts5ptTlamb}
\begin{aligned}
  &\vec{Q}=\frac{\vec{\tilde{Q}}_1}{4}+\frac{\vec{\tilde{Q}}_2}{4}+\frac{\vec{\tilde{Q}}_3}{4}+\frac{\vec{\tilde{Q}}_4}{4}\ ,\\
  &\vec{\tilde{Q}}_1\ =\ 
  4\begin{pmatrix}
    P^0_{\alpha}-\epsilon & \epsilon/3 & \epsilon/3 & \epsilon/3 \\
    P^1_{\alpha} & 0 & 0 & 0 \\
    \vdots & \vdots & \vdots & \vdots \\
    P^{D-1}_\alpha & 0 & 0 & 0\\
  \end{pmatrix},\quad
  \vec{\tilde{Q}}_2\ =\ 
  4\begin{pmatrix}
    \epsilon/3 & P^0_{\beta}-\epsilon & \epsilon/3 & \epsilon/3 \\
    0 & P^1_{\beta} & 0 & 0 \\
    \vdots & \vdots & \vdots & \vdots \\
    0 & P^{D-1}_\beta & 0 & 0 \\
  \end{pmatrix},\\[5pt]
  &\vec{\tilde{Q}}_3\ =\ 
  4\begin{pmatrix}
    \epsilon/3 & \epsilon/3 & P^0_{\gamma}-\epsilon & \epsilon/3 \\
    0 & 0 & P^1_{\gamma} & 0 \\
    \vdots & \vdots & \vdots & \vdots \\
    0 & 0 & P^{D-1}_{\gamma} & 0 \\
  \end{pmatrix},\quad
  \vec{\tilde{Q}}_4\ =\ 
  4\begin{pmatrix}
    \epsilon/3 & \epsilon/3 & \epsilon/3 & P^0_{\delta}-\epsilon\\
    0 & 0 & 0 & P^1_{\delta}\\
    \vdots & \vdots & \vdots & \vdots \\
    0 & 0 & 0 & P^{D-1}_\delta \\
  \end{pmatrix},
\end{aligned}
\end{equation}
with $0<\epsilon<\min\big\{P^0_r-\|\vec{P}_r\|,\ r=\alpha,\beta,\gamma,\delta\big\}$\footnote{In our notation, $P_r=(P^0_r,\vec{P}_r)$ and $\|\vec{P}_r\|=\sqrt{\sum_{i=1}^{D-1}(P^i_r)^2}$. Hereafter, $\|\cdot\|$ will denote the Euclidean norm of the $(D-1)$-dimensional spatial momenta sitting in its argument.}. Such a simple way of decomposing works for all $\vec{Q}\in\mathcal{C}$ due to the fact that the number of independent momenta for each $\vec{Q}$ in a primitive cone $\mathcal{C}$ listed in \eqref{eq:5ptTbases} is equal to the minimum number of $\text{Im }P_I$ that are needed to be specified in particular lightcones for the $\mathcal{C}$\footnote{The same is true for all the 32 primitive cones of 4-point functions and all the 6 primitive cones of 3-point functions. And, these primitive cones were obtained as the convex hulls of the corresponding LES cones \cite{BM2021}.}.

\section{Remaining primitive tubes}
\label{sec:remaintubes}

In this section, we consider the remaining 20 primitive tubes of 5-point functions. The respective primitive cones are given by \cite{AB1960}
\begin{equation}
\label{eq:probtubes}
\begin{aligned}
  &\mathcal{C}'^{+}_{ab}=-\mathcal{C}'^{-}_{ab}&=&\bigg\{\text{Im }p:\ \text{Im }(p_a+p_c),\ \text{Im }(p_a+p_d),\ \text{Im }(p_a+p_e),\\
  &&&\quad\quad\quad\quad\quad\text{Im }(p_b+p_c),\ \text{Im }(p_b+p_d),\ \text{Im }(p_b+p_e)\in V^+\bigg\}\ ,
\end{aligned}
\end{equation}
where $(abcde)=$ permutation of $(12345)$, and $\text{Im }p=(\text{Im }p_1,\dots,\text{Im }p_5)$ is linked by $\text{Im }p_1+\cdots+\text{Im }p_5=0$.

In the following subsection, for concreteness, we consider in detail the holomorphic extension of the LES cone $\bigcup_{\vec{\theta}}\mathcal{C}'^{+,\vec{\theta}}_{12}\equiv\tilde{\mathcal{C}}'^{+}_{12}$ inside the primitive cone $\mathcal{C}'^{+}_{12}$ (which is obtained when $(ab)=(12)$). Permuting $(abcde)$ simply generates the analysis for the other primitive cones listed in \eqref{eq:probtubes}.

\subsection{Extension inside $\mathcal{C}'^{+}_{12}$}
\label{sec:extensioninprobt}

Due to the external momentum conservation, we can choose any set of four linear combinations of external momenta which are linearly independent as our basis. Here, we choose the following set as our basis\footnote{We discuss its relation to another choice for a basis in appendix \ref{app:probtube1CT}.} in order to assign coordinates to the points in $\mathcal{C}'^{+}_{12}$
\begin{equation}
\begin{aligned}
  \bigg\{\text{Im }(p_1+p_3),\ \text{Im }(p_1+p_4),\ \text{Im }(p_1+p_5),\ \text{Im }(p_2+p_3)\bigg\}\ .
\end{aligned}
\end{equation}
With this choice, we can write
\begin{equation}
\label{eq:probtube1}
  \mathcal{C}'^{+}_{12}=\bigg\{\vec{Q}=(P_1,P_2,P_3,P_4):\ P_1,\ P_2,\ P_3,\ P_4,\ (P_2+P_4-P_1),\ (P_3+P_4-P_1)\in V^+\bigg\}\ .
\end{equation}
Now, any $\vec{Q}\in\mathcal{C}'^{+}_{12}$ can be explicitly written as a $D\times4$ matrix given by
\begin{equation}
\label{eq:genericpt}
  \vec{Q}\ =\ 
  \begin{pmatrix}
    P^0_{1} & P^0_{2} & P^0_{3} & P^0_{4} \\
    P^1_{1} & P^1_{2} & P^1_{3} & P^1_{4} \\
    \vdots & \vdots & \vdots & \vdots \\
    P^{D-1}_{1} & P^{D-1}_{2} & P^{D-1}_{3} & P^{D-1}_{4}
  \end{pmatrix},
\end{equation}
with conditions: $s_r>0,\ r=1,\dots,6$, where $s_r$ are defined by
\begin{equation}
\begin{aligned}
  &s_r = P^0_r-\|\vec{P}_r\|\ ,\quad r=1,\dots,4;\quad s_5 = P^0_2+P^0_4-P^0_1-\|\vec{P}_2+\vec{P}_4-\vec{P}_1\|\ ;\\
  &s_6 = P^0_3+P^0_4-P^0_1-\|\vec{P}_3+\vec{P}_4-\vec{P}_1\|\ .
\end{aligned}
\end{equation}

Also, define
\begin{equation}
\label{eq:defPij}
  P_{ij}\equiv P_i-P_j,\quad i\neq j,\quad i,j=1,\dots,4\ .
\end{equation}
As discussed in appendix \ref{app:probtube1CT}, for some points in $\mathcal{C}'^{+}_{12}$, few or all the $P_{ij}$ can be spacelike or lightlike. The two additional linear combinations of momenta $P_1,P_2,P_3,P_4$ needed to be in $V^+$ for $\mathcal{C}'^{+}_{12}$ can be obtained as linear combinations of $P_{14}$ and $P_2$ or $P_3$ respectively, as can be seen from \eqref{eq:probtube1}. Motivated by this observation, we classify the task of finding a convex decomposition of a $\vec{Q}\in\mathcal{C}'^{+}_{12}$ into points drawn from the LES cone $\tilde{\mathcal{C}}'^{+}_{12}$ as whether $P_{14}$ is timelike or lightlike or spacelike. At this stage, for each $\vec{Q}\in\mathcal{C}'^{+}_{12}$, we define a quantity $l_1$ by
\begin{equation}
\label{eq:defl1}
  l_1=\|\vec{P}_{14}\|-|P^0_{14}|\ .
\end{equation}
For a given $\vec{Q}\in\mathcal{C}'^{+}_{12}$, if $P_{14}$ is timelike or lightlike (i.e., $l_1\leq0$), in subsection \ref{sec:case1}, we show that $\vec{Q}\in\text{Ch}(\bigcup_{\vec{\theta}}\mathcal{C}'^{+,\vec{\theta}}_{12})$. In subsection \ref{sec:case2}, we examine the case when $P_{14}$ is spacelike (i.e., $l_1>0$), which is more subtle.

\subsubsection{$P_{14}$ is timelike or lightlike}
\label{sec:case1}

First, we consider the case when for a given $\vec{Q}\in\mathcal{C}'^{+}_{12}$, $P_{14}$ is timelike. In this case, $\vec{Q}$ can be written as a $D\times4$ matrix as in \eqref{eq:genericpt}, with conditions: $s_r>0,\ r=1,\dots,6$, and $l_1<0$. Now, there are two subcases, namely $P^0_{14}>0,P^0_{41}>0$. When $P^0_{14}>0$, $\vec{Q}$ can be decomposed as
\begin{equation}
\label{eq:Qdecompt1}
\begin{aligned}
  &\vec{Q}=\frac{\vec{\tilde{Q}}_1}{4}+\frac{\vec{\tilde{Q}}_2}{4}+\frac{\vec{\tilde{Q}}_3}{4}+\frac{\vec{\tilde{Q}}_4}{4}\ ,\\
  &\vec{\tilde{Q}}_1\ =\ 
  4\begin{pmatrix}
    P^0_{14}-\frac{\epsilon}{2} & P^0_{14}-\frac{\epsilon}{2} & P^0_{14}-\frac{\epsilon}{2} & \frac{\epsilon}{2} \\
    P^1_{14} & P^1_{14}  & P^1_{14} & 0 \\
    \vdots & \vdots & \vdots & \vdots \\
    P^{D-1}_{14} & P^{D-1}_{14} & P^{D-1}_{14} & 0\\
  \end{pmatrix},\quad
  \vec{\tilde{Q}}_2\ =\ 
  4\begin{pmatrix}
    \frac{\epsilon}{2} & P^0_{2}+P^0_{41}-\frac{3\epsilon}{4} & \frac{\epsilon}{2} & \frac{\epsilon}{4} \\
    0 & P^1_{2}+P^1_{41}  & 0 & 0 \\
    \vdots & \vdots & \vdots & \vdots \\
    0 & P^{D-1}_{2}+P^{D-1}_{41} & 0 & 0\\
  \end{pmatrix},\\[5pt]
  &\vec{\tilde{Q}}_3\ =\ 
  4\begin{pmatrix}
    \frac{\epsilon}{2} & \frac{\epsilon}{2} & P^0_{3}+P^0_{41}-\frac{3\epsilon}{4} & \frac{\epsilon}{4} \\
    0 & 0 & P^1_{3}+P^1_{41} & 0 \\
    \vdots & \vdots & \vdots & \vdots \\
    0 & 0 & P^{D-1}_{3}+P^{D-1}_{41} & 0 \\
  \end{pmatrix},\quad
  \vec{\tilde{Q}}_4\ =\ 
  4\begin{pmatrix}
    P^0_{4}-\frac{\epsilon}{2} & \frac{3\epsilon}{4} & \frac{3\epsilon}{4} & P^0_{4}-\epsilon \\
    P^1_{4} & 0 & 0 & P^1_{4} \\
    \vdots & \vdots & \vdots & \vdots \\
    P^{D-1}_{4} & 0 & 0 & P^{D-1}_{4} \\
  \end{pmatrix},
\end{aligned}
\end{equation}
with $0<\epsilon<\textrm{min}\{s_4,s_5,s_6,|l_1|\}$. For each $m=1,\dots,4$, all the four columns of $\vec{\tilde{Q}}_m$ belong to $V^+$. Furthermore, $\vec{\tilde{Q}}_m[2]+\vec{\tilde{Q}}_m[4]-\vec{\tilde{Q}}_m[1],\ \vec{\tilde{Q}}_m[3]+\vec{\tilde{Q}}_m[4]-\vec{\tilde{Q}}_m[1]\in V^+$, where $\vec{\tilde{Q}}_m[n]$ denotes the $n$-th column of $\vec{\tilde{Q}}_m ,\ n=1,\dots,4$. Also the four columns of $\vec{\tilde{Q}}_m$ lie on the same two dimensional Lorentzian plane, for each $m$. As a result, \eqref{eq:Qdecompt1} demonstrates that in the current subcase $\vec{Q}\in\text{Ch}(\bigcup_{\vec{\theta}}\mathcal{C}'^{+,\vec{\theta}}_{12})$.

At this stage, for later convenience, we introduce a timelike vector
\begin{equation}
  \bar{\epsilon}=\begin{pmatrix}\epsilon\\0\\\vdots\\0\end{pmatrix}\ ,
\end{equation}
using which and with appropriate $\epsilon$ mentioned above, \eqref{eq:Qdecompt1} can be abbreviated as
\begin{equation}
\label{eq:Qdecompshort}
\begin{aligned}
  (P_1,P_2,P_3,P_4)=&\ (P_{14}-\frac{\bar{\epsilon}}{2},P_{14}-\frac{\bar{\epsilon}}{2},P_{14}-\frac{\bar{\epsilon}}{2},\frac{\bar{\epsilon}}{2})+(\frac{\bar{\epsilon}}{2},P_2+P_{41}-\frac{3\bar{\epsilon}}{4},\frac{\bar{\epsilon}}{2},\frac{\bar{\epsilon}}{4})\\
  +&\ (\frac{\bar{\epsilon}}{2},\frac{\bar{\epsilon}}{2},P_3+P_{41}-\frac{3\bar{\epsilon}}{4},\frac{\bar{\epsilon}}{4})+(P_4-\frac{\bar{\epsilon}}{2},\frac{3\bar{\epsilon}}{4},\frac{3\bar{\epsilon}}{4},P_4-\bar{\epsilon})\ .
\end{aligned}
\end{equation}
In this and the next subsections, we will adopt abbreviations similar to that used in the above.

Now, we consider the subcase when for a given $\vec{Q}\in\mathcal{C}'^{+}_{12}$, $P_{41}\in V^+$. $\vec{Q}$ can be decomposed as\footnote{As previously indicated, we use abbreviations here, as in \eqref{eq:Qdecompshort}.}
\begin{equation}
\begin{aligned}
  (P_1,P_2,P_3,P_4)=&\ (P_1-\bar{\epsilon},\frac{\bar{\epsilon}}{2},\frac{\bar{\epsilon}}{2},P_1-\bar{\epsilon})+(\frac{\bar{\epsilon}}{2},P_2-\bar{\epsilon},\frac{\bar{\epsilon}}{4},\bar{\epsilon})\\
  +&\ (\frac{\bar{\epsilon}}{4},\frac{\bar{\epsilon}}{4},P_3-\bar{\epsilon},\bar{\epsilon})+(\frac{\bar{\epsilon}}{4},\frac{\bar{\epsilon}}{4},\frac{\bar{\epsilon}}{4},P_{41}-\bar{\epsilon})\ ,
\end{aligned}
\end{equation}
with $0<\epsilon<\textrm{min}\{s_1,s_2,s_3,|l_1|\}$. Hence, $\vec{Q}\in\text{Ch}(\bigcup_{\vec{\theta}}\mathcal{C}'^{+,\vec{\theta}}_{12})$, in this subcase as well.

Secondly, we consider the case when for a given $\vec{Q}\in\mathcal{C}'^{+}_{12}$, $P_{14}$ is lightlike. In this case, $\vec{Q}$ can be written as a $D\times4$ matrix as in \eqref{eq:genericpt}, with conditions: $s_r>0,\ r=1,\dots,6$, and $l_1=0$. Now, there are two subcases, namely $P^0_{14}\geq0,P^0_{41}>0$\footnote{Since $l_1=0$, the equality $P^0_{14}=0$ implies that $P_1=P_4$.}. When $P^0_{14}\geq0$, $\vec{Q}$ can be decomposed as
\begin{equation}
\begin{aligned}
  (P_1,P_2,P_3,P_4)=&\ (P_{14}+\frac{\bar{\epsilon}}{3},P_{14}+\frac{\bar{\epsilon}}{3},P_{14}+\frac{\bar{\epsilon}}{3},\frac{\bar{\epsilon}}{3})+(\frac{\bar{\epsilon}}{3},P_2+P_{41}-\bar{\epsilon},\frac{\bar{\epsilon}}{3},\frac{\bar{\epsilon}}{3})\\
  +&\ (\frac{\bar{\epsilon}}{3},\frac{\bar{\epsilon}}{3},P_3+P_{41}-\bar{\epsilon},\frac{\bar{\epsilon}}{3})+(P_4-\bar{\epsilon},\frac{\bar{\epsilon}}{3},\frac{\bar{\epsilon}}{3},P_4-\bar{\epsilon})\ ,
\end{aligned}
\end{equation}
with $0<\epsilon<\textrm{min}\{s_4,s_5,s_6\}$.  When $P^0_{41}>0$, $\vec{Q}$ can be decomposed as
\begin{equation}
\begin{aligned}
  (P_1,P_2,P_3,P_4)=&\ (P_1-\bar{\epsilon},\frac{\bar{\epsilon}}{3},\frac{\bar{\epsilon}}{3},P_1-\bar{\epsilon})+(\frac{\bar{\epsilon}}{3},P_2-\bar{\epsilon},\frac{\bar{\epsilon}}{3},\frac{\bar{\epsilon}}{3})\\
  +&\ (\frac{\bar{\epsilon}}{3},\frac{\bar{\epsilon}}{3},P_3-\bar{\epsilon},\frac{\bar{\epsilon}}{3})+(\frac{\bar{\epsilon}}{3},\frac{\bar{\epsilon}}{3},\frac{\bar{\epsilon}}{3},P_{41}+\frac{\bar{\epsilon}}{3})\ ,
\end{aligned}
\end{equation}
with $0<\epsilon<\textrm{min}\{s_1,s_2,s_3\}$. Therefore, when $P_{14}$ is lightlike for a $\vec{Q}\in\mathcal{C}'^{+}_{12}$, we obtain $\vec{Q}\in\text{Ch}(\bigcup_{\vec{\theta}}\mathcal{C}'^{+,\vec{\theta}}_{12})$.

\subsubsection{$P_{14}$ is spacelike}
\label{sec:case2}

In this subsection, we consider the case when for a given $\vec{Q}\in\mathcal{C}'^{+}_{12}$, $P_{14}$ is spacelike. A real Lorentz transformation maps a point in a primitive tube to another point in the same tube. To study the analyticity at a point within a primitive tube, we can do the analysis in any Lorentz frame without losing generality\footnote{This is supported by Cor1 as given in \eqref{eq:LEScor}.}. Thus, we restrict ourselves to $\vec{Q}\in\mathcal{C}'^{+}_{12}$ for which $P_{14}$ is spacelike with $P^0_{14}=0$ i.e., $P^0_1=P^0_4$, in a given Lorentz frame. Therefore, from \eqref{eq:defl1} we get
\begin{equation}
  l_1=\|\vec{P}_{14}\|\ .
\end{equation}
In this case, $\vec{Q}$ can be written as a $D\times4$ matrix as in \eqref{eq:genericpt}, with conditions: $s_r>0,\ r=1,\dots,6$, and $l_1>0$. Also, we have
\begin{equation}
\label{eq:probtube1slconds}
\begin{aligned}
  s_1\leq s_4+l_1,\ s_4\leq s_1+l_1\ ,\\
  s_2\leq s_5+l_1,\ s_5\leq s_2+l_1\ ,\\
  s_3\leq s_6+l_1,\ s_6\leq s_3+l_1\ .\\
\end{aligned}
\end{equation}
For points $\vec{Q}\in\mathcal{C}'^{+}_{12}$ in the present consideration, the allowed space in parameters $s_1,\dots,s_6,l_1$ will be determined by the above inequalities. We denote this allowed space by $\mathcal{S}$. As we will see below, different ways of decomposing $\vec{Q}$ into points taken from the LES cone $\tilde{\mathcal{C}}'^{+}_{12}$ (whose convex combination represents the $\vec{Q}$) work in different sections of $\mathcal{S}$. We use a specific algorithm (as outlined in section \ref{sec:intro} for a more generic case) to provide such types of convex decomposition. In the present case, the algorithm analytically obtains types of convex decomposition involving only four terms along with the region of validity for each type. We select any four momenta from the set\footnote{In appendix \ref{app:algo2} we discuss how this set may be modified to obtain other possible types of decomposition.}
\begin{equation}
\label{eq:algostep2}
  \bigg\{P_i,\ P_2+P_4-P_1,\ P_3+P_4-P_1,\ P_{ij},\ i\neq j,\ i,j=1,\dots,4\bigg\}\ ,
\end{equation}
such that the inverse transformation from them to $\{P_1,P_2,P_3,P_4\}$ exists. This allows to write each $P_i,\ i=1,\dots,4$ as a linear combination of the chosen four momenta. Using this we decompose the point $(P_1,P_2,P_3,P_4)$ into four terms, with all the momenta in each term lying on a two dimensional Lorentzian plane which will be the plane of one of the chosen momenta. With all consistent slections of the initial set of four momenta from \eqref{eq:algostep2}, we have a list of decompositions into four terms that involve only the chosen four momenta. We accept only those types of decomposition from this list for which, improving only the temporal components of the momenta in the terms, all the four terms can be brought inside $\tilde{\mathcal{C}}'^{+}_{12}$ maintaining that the terms add up to $(P_1,P_2,P_3,P_4)$. We call the prescription to do this as a ``consistent $\epsilon$-prescription''. 

Let us denote the selected four momenta by $P_{i_1}, P_{i_2}, P_{i_3}, P_{i_4}$ and the inverse transformation matrix from these four to $P_1, P_2, P_3, P_4$ by $\mathbb{A}$. Now using $\mathbb{A}^{T}$ we get,
\begin{equation}
\begin{aligned}
(P_{1},\ P_{2},\ P_{3},\ P_{4})&=(\mathbb{A}^{T}_{11}P_{i_1}, \ \mathbb{A}^{T}_{12}P_{i_1}, \ \mathbb{A}^{T}_{13}P_{i_1}, \ \mathbb{A}^{T}_{14}P_{i_1})+(\mathbb{A}^{T}_{21}P_{i_2}, \ \mathbb{A}^{T}_{22}P_{i_2}, \ \mathbb{A}^{T}_{23}P_{i_2}, \ \mathbb{A}^{T}_{24}P_{i_2})\\
&+(\mathbb{A}^{T}_{31}P_{i_3}, \ \mathbb{A}^{T}_{32}P_{i_3}, \ \mathbb{A}^{T}_{33}P_{i_3}, \ \mathbb{A}^{T}_{34}P_{i_4})+(\mathbb{A}^{T}_{41}P_{i_4}, \ \mathbb{A}^{T}_{42}P_{i_4}, \ \mathbb{A}^{T}_{43}P_{i_1}, \ \mathbb{A}^{T}_{44}P_{i_4})\ .
\end{aligned}
\label{eq:matrixcond}
\end{equation}
For the aforementioned search of valid types of convex decomposition the conditions imposed are equivalent to the following.
For each term on the right hand side of \eqref{eq:matrixcond} we must have,\\
$\mathbb{A}^T_{j1},\ \mathbb{A}^T_{j2},\ \mathbb{A}^T_{j3},\ \mathbb{A}^T_{j4},\ \mathbb{A}^T_{j2}+\mathbb{A}^T_{j4}-\mathbb{A}^T_{j1}\ \ \text{and}\ \ \mathbb{A}^T_{j2}+\mathbb{A}^T_{j4}-\mathbb{A}^T_{j1}$
\begin{eqnarray}
 && \text{non-negative if $P_{i_j}\in\lbrace P_1,\dots,P_4,P_2+P_{41}, P_3+P_{41}\rbrace$}\ ;\nonumber\\
&& \text{no restriction if $P_{i_j}=P_{14}$}\ ;\nonumber\\
&& \text{of the same sign or 0 otherwise}\ .
\end{eqnarray}

In the final list, there are 8 types of convex decomposition that involve only one $P_{ij}$, which is $P_{14}$ in each instance. There are 53 types of convex decomposition involving two $P_{ij}$ and 68 types involving three $P_{ij}$, which exhausts all possibilities. In appendix \ref{app:Qdecomp}, we list all types of decomposition obtained by the above algorithm, and give the full region of validity for only a few examples. In current subsection, we call them whenever required and explicitly present consistent $\epsilon$-prescriptions for those who are called.

Inside the aforementioned allowed space $\mathcal{S}$, firstly we consider the eight subregions $U_r,\ r=1,2,\dots,8$ given by
\begin{equation}
\begin{aligned}
  &U_1:\ s_1,s_2,s_3>l_1>0,\quad U_2:\ s_1,s_3,s_5>l_1>0,\quad U_3:\ s_1,s_2,s_6>l_1>0\ ,\\
  &U_4:\ s_1,s_5,s_6>l_1>0,\quad U_5:\ s_2,s_3,s_4>l_1>0,\quad U_6:\ s_3,s_4,s_5>l_1>0\ ,\\
  &U_7:\ s_2,s_4,s_6>l_1>0,\quad U_8:\ s_4,s_5,s_6>l_1>0\ .
\end{aligned}
\end{equation}
For each $r=1,2,\dots,8$, any $\vec{Q}\in U_r$ can be written as a convex combination of four points taken from the LES cone $\tilde{\mathcal{C}}'^{+}_{12}$, respectively using the types of decomposition given in table \ref{tab:decomp1Pij} of appendix \ref{app:Qdecomp}. As an example, in the following we explicitly give the type of convex decomposition that works for all $\vec{Q}\in U_1$.
\begin{equation}
\label{eq:QdecompU1expl}
\begin{aligned}
  (P_1,P_2,P_3,P_4)=&\ (P_1-\bar{\epsilon},\ \frac{1-k}{2}\bar{\epsilon},\ \frac{1-k}{2}\bar{\epsilon},\ P_1-\bar{\epsilon})+(\frac{1-k}{2}\bar{\epsilon},\ P_2-\bar{\epsilon},\ \frac{1-k}{2}\bar{\epsilon},\ \frac{1-k}{2}\bar{\epsilon})\\
  +&\ (\frac{1-k}{2}\bar{\epsilon},\ \frac{1-k}{2}\bar{\epsilon},\ P_3-\bar{\epsilon},\ \frac{1-k}{2}\bar{\epsilon})+(k\bar{\epsilon},\ k\bar{\epsilon},\ k\bar{\epsilon},\ P_{41}+ k\bar{\epsilon})\ ,
\end{aligned}
\end{equation}
with any $k$ chosen as $l_1/s<k<1$, and any $\epsilon$ chosen as $l_1/k<\epsilon<s,\ s=\textrm{min}\{s_1,s_2,s_3\}$.

Now, inside $\mathcal{S}$, the complement of $\cup_{r=1}^8U_r$ is given by
\begin{equation}
\label{eq:complementreg}
  (s_1,s_4\leq l_1)\cup(s_2,s_5\leq l_1)\cup(s_3,s_6\leq l_1)\ ,
\end{equation}
subject to $s_1,\dots,s_6,l_1>0$ and \eqref{eq:probtube1slconds}. With \eqref{eq:complementreg}, $P_{ij}$ (other than $P_{14}$) can be timelike, lightlike, or spacelike. Thus, we introduce the following parameters
\begin{equation}
\begin{aligned}
  &l_2=\|\vec{P}_{12}\|-|P^0_{12}|,\ l_3=\|\vec{P}_{13}\|-|P^0_{13}|,\ l_4=\|\vec{P}_{23}\|-|P^0_{23}|\ ,\\
  &l_5=\|\vec{P}_{24}\|-|P^0_{24}|,\ l_6=\|\vec{P}_{34}\|-|P^0_{34}|\ .
\end{aligned}
\end{equation}
Clearly, when any $l_r\ (r\neq1)$ is greater than or equal to zero, the corresponding $P_{ij}$ is spacelike or lightlike, respectively. And, if $l_r$ is less than zero, the corresponding $P_{ij}$ is timelike. Various parts of the above region \eqref{eq:complementreg} are respectively covered by the types of convex decomposition involving two or three $P_{ij}$ (i.e., table \ref{tab:decomp2Pij} and table \ref{tab:decomp3Pij} of appendix \ref{app:Qdecomp}). Below, we consider a few such types in order to discuss the associated $\epsilon$-prescriptions. Then, the question of whether a point exists inside $\mathcal{S}$ which cannot be decomposed in any of the ways that are obtained using our algorithm will only be answered in section \ref{sec:numerics}.

Now, we discuss a consistent $\epsilon$-prescription for the decomposition no. 43 of table \ref{tab:decomp2Pij}, as an example of the types of convex decomposition involving two $P_{ij}$. The type of convex decomposition in consideration is valid for all points $\vec{Q}$ with $P^0_{21},P^0_{31}>0$ and $s_1,s_4>\max\{l_2,l_3\}$. Using notations $l=\max\{l_2,l_3\},s=\min\{s_1,s_4\}$, for this region we have $s>l,s>0$. The type of decomposition is explicitly given by
\begin{equation}
\begin{aligned}
  (P_1,P_2,P_3,P_4)&=(P_1-\bar{\epsilon},P_1-\bar{\epsilon},P_1-\bar{\epsilon},\frac{1-k}{2}\bar{\epsilon})+(\frac{1-k}{2}\bar{\epsilon},\frac{1-k}{2}\bar{\epsilon},\frac{1-k}{2}\bar{\epsilon},P_4-\bar{\epsilon})\\
  &+(\frac{1-k}{2}\bar{\epsilon},P_{21}+k\bar{\epsilon},\frac{1-k}{2}\bar{\epsilon},\frac{1-k}{2}\bar{\epsilon})+(k\bar{\epsilon},\frac{1-k}{2}\bar{\epsilon},P_{31}+k\bar{\epsilon},k\bar{\epsilon})\ ,
\end{aligned}
\end{equation}
with $l/s<k<1,\ l/k<\epsilon<s$ when $l>0$, and $0<k<1,\ 0<\epsilon<s$ when $l\leq0$.

Now, we discuss a consistent $\epsilon$-prescription for the decomposition no. 1 of table \ref{tab:decomp3Pij}, as an example of the types of convex decomposition involving three $P_{ij}$. The type of convex decomposition in consideration is valid for all points $\vec{Q}$ with $P^0_{21},P^0_{31}>0$ and $s_1>l_1,s_1>\max\{l_1+l_2,l_1+l_3\}$. Using notations $l=\max\{l_2,l_3\},l'=\max\{l_1+l_2,l_1+l_3\}$, for this region we have $s_1>l_1>0,s_1>l'$. The type of decomposition is explicitly given by
\begin{equation}
\begin{aligned}
  (P_1,P_2,P_3,P_4)&=(P_1-\bar{\epsilon},P_1-\bar{\epsilon},P_1-\bar{\epsilon},P_1-\bar{\epsilon})\\
  &+\left((1-k'-k)\bar{\epsilon},P_{21}+k'\bar{\epsilon},(1-k'-k)\bar{\epsilon},(1-k'-k)\bar{\epsilon}\right)\\
  &+\left(k'\bar{\epsilon},(1-k'-k)\bar{\epsilon},P_{31}+k'\bar{\epsilon},k'\bar{\epsilon}\right)+\left(k\bar{\epsilon},k\bar{\epsilon},k\bar{\epsilon},P_{41}+k\bar{\epsilon}\right)\ ,
\end{aligned}
\end{equation}
with $\{k,k',\epsilon\}$ satisfying $k,k'>0,\ k+k'<1,\ 0<\epsilon<s_1,\ k\epsilon>l_1,\ k'\epsilon>l$. Now, $\{k,k',\epsilon\}$ always exists only if $s_1>l_1>0,s_1>l_1+l\ (=l')$ (providing conditions to obtain the region of validity).

\section{Numerical analysis for $\mathcal{C}'^{+}_{12}$ }
\label{sec:numerics}

In this section, we numerically generate distinct points inside the primitive cone $\mathcal{C}'^{+}_{12}$ taking $P_1,\dots,P_4$ in 3 space-time dimensions. There are a total of 53 types of convex decomposition involving two $P_{ij}$ and 68 types involving three $P_{ij}$, obtained using our algorithm as mentioned in section \ref{sec:case2}. Among these only 33 are such, that do not involve $P_{14}$. Any point in the region\footnote{Clearly, we have $\mathcal{S}'\subsetneq\mathcal{S}$.}
\begin{equation}
  \mathcal{S}'=(s_1,s_4\leq l_1)\cap(s_2,s_5\leq l_1)\cap(s_3,s_6\leq l_1)
\end{equation}
(subject to $s_1,\dots,s_6,l_1>0$ and \eqref{eq:probtube1slconds}) if at all covered, has to be covered using these 33 types of convex decomposition\footnote{Since the types of convex decomposition that involve $P_{14}$ must have $l_1$ smaller than at least one $s_r$ in their region of validity.}. If we find a point that is not covered by them, we can surely say that it will not be covered by the types of convex decomposition derived by our algorithm. We find that indeed there are many such points and below we give numerical results in support of this statement. We present two explicit examples from the leftover points.

First, we have to specify the total number independent variables required to generate different configurations\footnote{A configuration is identified by the orientation of the spatial parts (vector) of the momenta $P_1,...,P_4$ and also whether $P_2^0\geq P_1^0, P_4^0\ /\ P_2^0< P_1^0, P_4^0$ and $P_3^0\geq P_1^0, P_4^0\ /\ P_3^0< P_1^0, P_4^0$.}.
\begin{enumerate}
  \item{Since we are concerned with the case where $P_{14}$ is space-like, we can always go to a Lorentz frame where the temporal component of it vanishes, i.e. in this frame we have, $P_1^0=P_4^0$. This requires a Lorentz boost.}
  \item{Now that we have fixed some boosted frame we still have some freedom regarding pure rotations. We use it to set at least one of the vector ($P_1$ in our case) to reside on the $p^0-p^1$ plane, i.e. after the rotation $P_1$ takes the form
\begin{equation}
  P_1=\begin{pmatrix}
  P^0_1\\
  P^1_1\\
  0
  \end{pmatrix}\ .
\end{equation}
}
  \item{Finally, we note that the overall scaling of $P_1,\dots,P_4$ is not important since the types of convex decomposition are independent of the overall scaling, i.e. if a type works for a particular configuration, the same type works for the configuration obtained by an overall scaling\footnote{This is owing to the fact that for every cone $A$, $p\in A\implies\alpha p\in A\ \forall\alpha>0$, which is also true for primitive cones.}. So, we can set $P_1^0=P_4^0=1$ without loss of generality. Thus, we are left with a total of 9 independent variables instead of 12.
\begin{equation}
  (P_1,\ P_2,\ P_3,\ P_4)=
  \begin{pmatrix}
  1 & P_2^0 & P_3^0 & 1 \\
  P_1^1 & P_2^1 & P_3^1 & P_4^1 \\
  0 & P_2^2 & P_3^2 & P_4^2
  \end{pmatrix}.
\end{equation}
}
\end{enumerate}
Note that the values of $P_1^1,P_2^0,P_3^0,P_2^1,P_2^2,P_3^1,P_3^2,P_4^1, P_4^2$ are constrained such that we have $s_1, s_2,\dots,s_6>0$.\\

\subsection{Region covered inside $\mathcal{S}'$}

We give the values of $P_2^0$ and $P_3^0$ as input and the algorithm generates configurations with $|\vec{P}_i|$ taking values from $0$ to $0.98 P_i^0$ for all $i=1,\dots,4$ in $m$ steps, and $P_i^1=|\vec{P}_i|\cos\theta_i,\ P_i^2=|\vec{P}_i|\sin\theta_i$, with $\theta_1=0$ while $\theta_2,\theta_3,\theta_4$ take values $0$ to $2\pi$ in $n$ steps (we approximate $\pi\approx 3.14159$). This only guarantees the positivity of $s_1,s_2,s_3$ and $s_4$. Among all the configurations generated this way we keep only the distinct configurations for which $s_5>0$ and $s_6>0$. The number of points in the collection after this step gives the total number of points inside the primitive cone $\mathcal{C}^{\prime(5) +}_{12}$. From this collection, we identify the points in $\mathcal{S}'$ and the region inside $\mathcal{S}'$ which is covered by the 33 types of convex decomposition which do not involve $P_{14}$. Below, we present our numerical results in table \ref{tab:result} for $m=11,n=13$.
\begin{table}[ht]
\caption{Points covered inside $\mathcal{S}'$; $m=11,n=13$ }
\centering
  \begin{tabular}{c c c c c}
  \hline\hline
  $P_2^0$ & $P_3^0$ & Total points inside $\mathcal{S}'$ & Covered points & Percentage \\
  \hline
  0.4 & 0.5 & 97,736 & 10,230 & 10.467\% \\
  0.8 & 1.2 & 9,19,956 & 1,48,352 & 16.126\% \\
  1 & 1 & 14,21,204 & 2,25,852 & 15.8916\% \\
  2 & 3 & 3,95,962 & 69,338 & 17.5113\% \\ [1ex]
  \hline
  \end{tabular}
\label{tab:result}
\end{table}

\subsection{Points outside the covered region}

As one can see, there are many points in $\mathcal{S}'$ outside the covered region and these cannot be decomposed in any of the ways obtained using the algorithm described in section \ref{sec:case2}. We give two examples of such points. It may be still possible to decompose these points, but for the check, we will need a separate algorithm to analytically generate new types of convex decomposition (see appendix \ref{app:algo2}) or numerical techniques. We will not consider those in the present section. Consider,
\begin{equation}
  P_1=P_2=
  \begin{pmatrix}
    1 \\
    0.098 \\
    0
  \end{pmatrix},\ \
  P_3=
  \begin{pmatrix}
    1 \\
    -0.098 \\
    0
  \end{pmatrix},\ \
  P_4=
  \begin{pmatrix}
    1 \\
    -0.441 \\
    0.7638
  \end{pmatrix}.
\end{equation}
We can compute the following values,
\begin{equation}
\begin{aligned}
  &s_1=s_2= s_3=0.902,\ s_4=s_5=0.118,\ s_6=0.0054\ ,\\
  &l_1=0.935,\ l_2=0,\ l_3=l_4=0.196,\ l_5=0.935,\ l_6=0.837\ .
\end{aligned}
\end{equation}

We now turn to a more generic point that is to say a case where no two of the vectors $P_1,\dots,P_4$ are equal to each other.
\begin{equation}
  P_1=
  \begin{pmatrix}
    1 \\
    0.098 \\
    0
  \end{pmatrix},\ \ 
  P_2=
  \begin{pmatrix}
    1 \\
    0.08487 \\
    0.049
  \end{pmatrix},\ \ 
  P_3=
  \begin{pmatrix}
    1 \\
    -0.049 \\
    0.08487
  \end{pmatrix},\ \ 
  P_4=
  \begin{pmatrix}
    1 \\
    -0.7638 \\
    -0.441
  \end{pmatrix}.
\end{equation}
We can compute the following values,
\begin{equation}
\begin{aligned}
  &s_1=s_2=s_3=0.902,\ ,\ s_4=0.11803,\ s_5=0.1298,\ s_6=0.02\ ,\\
  &l_1=0.968,\ l_2=0.05,\ l_3=0.1697,\ l_4=0.1386,\ l_5=0.98,\ l_6=0.8874\ .
\end{aligned}
\end{equation}
Note that in both of the above examples, we have $0<s_1,s_2,\dots,s_6<l_1$. Hence, they lie inside $\mathcal{S}'$.

\section{Discussions}
\label{sec:discussion}

In this article, we have studied the analytic extension of the LES tube inside each of the remaining 20 primitive tubes of off-shell 5-point loop amplitudes of closed superstring field theory. Let us summarize our main results in the following.
\begin{enumerate}
  \item{We have obtained types of convex decomposition by the inversion of some transformation matrix along with a consistent $\epsilon$-prescription that determines their region of validity. Such types of decomposition cover different sections of a primitive tube.}
  \item{We have shown that these 129 types of convex decomposition into four points drawn from the LES tube, cover only a portion of the primitive tube. We present explicit examples of the leftover points found employing numerical techniques.}
  \item{We have also briefly discussed ways to improve our algorithm.}
\end{enumerate}
Because we were confined to a specific algorithm, we may not have determined the whole convex hull of the LES tube. Therefore, points that are not covered by the 129 types of convex decomposition obtained by the preceding algorithm may still be decomposed into terms drawn from the LES tube. Appendix \ref{app:algo2} discusses improvements of our algorithm, although we leave the detailed analysis for future work since the computations become very tedious as the number of types of decomposition keep increasing.

As pointed in \cite{Ruelle1961}, similar to the aforementioned 20 tubes of 5-point functions, there can be a few primitive tubes for higher point functions where the number of independent external momenta (after accounting for external momentum conservation) is less than the minimum number of specific combinations of external momenta whose imaginary parts are required to be in specific light cones. For those cases as well, the general strategy and techniques of this article will apply.

Another potential future direction will be to develop an automated algorithm which gives a convex decomposition taking the leftover points numerically or at least confirm whether or not a decomposition is at all possible for the said points. 
As per our current understanding, there are no fundamental obstructions to deal with the remaining tubes. Therefore, rather dealing with case by case for all $n$, this direction is appealing to carry out the proof using a computer for arbitrary $n$-point functions.

\section*{Acknowledgements}

This work is supported in part by grants from the Infosys Foundation to CMI and HRI.

\appendix

\section{$\mathcal{C}'^{+}_{12}$ with a different basis}
\label{app:probtube1CT}

Due to the link $\text{Im }p_1+\cdots+\text{Im }p_5=0$ for points in $\mathcal{C}'^{+}_{12}$, we can choose any set of four linear combinations of $\text{Im }p_1,\dots,\text{Im }p_5$ which are linearly independent as our set of basis vectors, to assign coordinates to its points. Coordinates $(\text{Im }p_1,\text{Im }p_2,\text{Im }p_3,\text{Im }p_4)$ are assigned to $\vec{Q}\in\mathcal{C}'^{+}_{12}$, in the following basis
\begin{equation}
  \bigg\{\text{Im }p_1,\ \text{Im }p_2,\ \text{Im }p_3,\ \text{Im }p_4 \bigg\}\ .
\end{equation}
The coordinate transformation from $(P_1,P_2,P_3,P_4)$ (as given in \eqref{eq:probtube1}) to the aforementioned ones is an invertible linear map, given by
\begin{equation}
\begin{aligned}
  &\text{Im }p_1=P_2+P_3+P_4,\quad\text{Im }p_2=-P_1+P_2+P_3+2P_4\ ,\\
  &\text{Im }p_3=P_1-P_2-P_3-P_4,\quad\text{Im }p_4=-P_3-P_4\ ,
\end{aligned}
\end{equation}
with inversion formulae given by
\begin{equation}
\begin{aligned}
  &P_1=\text{Im }(p_1+p_3),\quad P_2=\text{Im }(p_1+p_4)\ ,\\
  &P_3=\text{Im }(p_1+p_5)=-\text{Im }(p_2+p_3+p_4),\quad P_4=\text{Im }(p_2+p_3)\ .
\end{aligned}
\end{equation}
From the definition of $\mathcal{C}'^{+}_{12}$ as in \eqref{eq:probtubes}, coordinates $(\text{Im }p_1,\text{Im }p_2,\text{Im }p_3,\text{Im }p_4)$ satisfy\footnote{All additional $\text{Im } P_I$ involving $\text{Im } p_5$ that are not listed in \eqref{eq:probtube1det} also belong to particular lightcones that can be easily determined using $\text{Im }p_1+\cdots+\text{Im }p_5=0$.}
\begin{equation}
\label{eq:probtube1det}
\begin{aligned}
  &\text{Im }p_1,\ \text{Im }p_2,\ -\text{Im }p_3,\ -\text{Im }p_4,\ \text{Im }(p_1 + p_2),\ \text{Im }(p_1 + p_3),\ \text{Im }(p_1 + p_4),\ \text{Im }(p_2 + p_3)\ ,\\
  &\text{Im }(p_2 + p_4),\ -\text{Im }(p_3 + p_4),\ \text{Im }(p_1 + p_2 + p_3),\ \text{Im }(p_1 + p_2 + p_4),\ -\text{Im }(p_1 + p_3 + p_4)\ ,\\
  &-\text{Im }(p_2 + p_3 + p_4),\ \text{Im }(p_1 +p_2 +p_3 +p_4) \in V^+\ .
\end{aligned}
\end{equation}
Now, $P_{ij}$ as defined in \eqref{eq:defPij} can be written as
\begin{equation}
\begin{aligned}
  &P_{12}=\text{Im }(p_3-p_4),\quad P_{13}=\text{Im }(p_1+p_2+2p_3+p_4)\ ,\\
  &P_{14}=\text{Im }(p_1-p_2),\quad P_{23}=\text{Im }(p_1 + p_2 + p_3 + 2p_4)\ ,\\
  &P_{24}=\text{Im }(p_1 - p_2 - p_3 + p_4),\quad P_{34}=-\text{Im }(2 p_2 + 2 p_3 +p_4)\ .
\end{aligned}
\end{equation}
To establish if any $P_{ij}$ belongs to a certain lightcone, \eqref{eq:probtube1det} is no longer adequate. In fact, there are points in $\mathcal{C}'^{+}_{12}$, for which few or all the $P_{ij}$ are spacelike or lightlike. Several instances of these types are already included in the numerically generated points in section \ref{sec:numerics}.

\section{Decompositions derived from the algorithm}
\label{app:Qdecomp}

In this appendix, we provide all the types of convex decomposition for $\vec{Q}\in\mathcal{C}'^{+}_{12}$ with spacelike $P_{14}$ derived from the algorithm as described in section \ref{sec:case2} (more generally in section \ref{sec:intro}). In a given Lorentz frame, confining ourselves to $\vec{Q}$ for which the temporal component of $P_{14}$ is zero will suffice. To execute the algorithm, we need to select four momenta from the set $\{P_i,\ P_2+P_4-P_1,\ P_3+P_4-P_1,\ P_{ij},\ i\neq j,\ i,j=1,\dots,4\}$. Following the algorithm, the final outcomes are types of convex decomposition into four terms (involving the selected momenta) drawn from the LES cone $\tilde{\mathcal{C}}'^{+}_{12}$. We do not obtain any types of convex decomposition involving all four from $\{P_1,P_2,P_3,P_4,P_2+P_4-P_1,P_3+P_4-P_1\}$ or involving four $P_{ij}$. There are 8 types of convex decomposition that involve only one $P_{ij}$, which is $P_{14}$ in each case. There are 53 types of convex decomposition involving two $P_{ij}$ and 68 types involving three $P_{ij}$. Only 33 of them do not involve $P_{14}$. Below, we present all the types (prior to the modification of temporal components) by grouping them according to how many $P_{ij}$ are involved. We present the region of validity for only a few of them. And, we detail the modification method for one type in order to highlight the general technique for obtaining their region of validity.\\

$\mathcal{B}.1$ \textbf{\textit{Decompositions that involve a single $P_{ij}$}}\\

\noindent In table \ref{tab:decomp1Pij}, we provide the 8 types of decomposition that involve a single $P_{ij}$ along with their region of validity. Then, for one of them, we elaborate the procedure to determine the corresponding region of validity.
\begin{table}[h!]
\begin{center}
  \begin{tabular}{|l|l|l|}
    \hline
    & $(P_1,P_2,P_3,P_4)=$ & Region of validity\\
    \hline
    1 & $(P_1,0,0,P_1)+(0,P_2,0,0)+(0,0,P_3,0)+(0,0,0,P_{41})$ & $s_1,s_2,s_3>l_1>0$\\
    \hline
    2 & $(P_1,0,0,P_1)+(0,P_2+P_{41},0,0)+(0,0,P_3,0)+(0,P_{14},0,P_{41})$ & $s_1,s_3,s_5>l_1>0$\\
    \hline
    3 & $(P_1,0,0,P_1)+(0,P_2,0,0)+(0,0,P_3+P_{41},0)+(0,0,P_{14},P_{41})$ & $s_1,s_2,s_6>l_1>0$\\
    \hline
    4 & $(P_1,0,0,P_1)+(0,P_2+P_{41},0,0)+(0,0,P_3+P_{41},0)+(0,P_{14},P_{14},P_{41})$ & $s_1,s_5,s_6>l_1>0$\\
    \hline
    5 & $(0,P_2,0,0)+(0,0,P_3,0)+(P_4,0,0,P_4)+(P_{14},0,0,0)$ & $s_2,s_3,s_4>l_1>0$\\
    \hline
    6 & $(0,P_2+P_{41},0,0)+(0,0,P_3,0)+(P_4,0,0,P_4)+(P_{14},P_{14},0,0)$ & $s_3,s_4,s_5>l_1>0$\\
    \hline
    7 & $(0,P_2,0,0)+(0,0,P_3+P_{41},0)+(P_4,0,0,P_4)+(P_{14},0,0,0)$ & $s_2,s_4,s_6>l_1>0$\\
    \hline
    8 & $(0,P_2+P_{41},0,0)+(0,0,P_3+P_{41},0)+(P_4,0,0,P_4)+(P_{14},P_{14},P_{14},0)$ & $s_4,s_5,s_6>l_1>0$\\
    \hline
  \end{tabular}
\end{center}
\caption{Types of decomposition that involve a single $P_{ij}$}
\label{tab:decomp1Pij}
\end{table}

Now, we consider the first type in detail:
\begin{equation}
\label{eq:QdecompU1}
\begin{aligned}
  (P_1,P_2,P_3,P_4)=(P_1,0,0,P_1)+(0,P_2,0,0)+(0,0,P_3,0)+(0,0,0,P_{41})\ .
\end{aligned}
\end{equation}
The above type of decomposition can be improved in the following way, in order to bring each of the four terms inside $\tilde{\mathcal{C}}'^{+}_{12}$.
\begin{equation}
\begin{aligned}
  &(P_1,P_2,P_3,P_4)=(P_1-\bar{\alpha}_1, \bar{\beta_1}, \bar{\gamma}_1,P_1-\bar{\delta}_1)+(\bar{\alpha}_2,P_2-\bar{\beta}_2, \bar{\gamma}_2, \bar{\delta}_2)\\
  &\qquad\qquad\qquad\ \ \ +(\bar{\alpha}_3, \bar{\beta}_3,P_3-\bar{\gamma}_3, \bar{\delta}_3)+(\bar{\alpha}_4, \bar{\beta}_4, \bar{\gamma}_4,P_{41}+ \bar{\delta}_4)\ ,\\
  &\bar{\alpha}_r=\begin{pmatrix}\alpha_r\\0\\\vdots\\0\end{pmatrix},\ \bar{\beta}_r=\begin{pmatrix}\beta_r\\0\\\vdots\\0\end{pmatrix},\ \bar{\gamma}_r=\begin{pmatrix}\gamma_r\\0\\\vdots\\0\end{pmatrix},\ \bar{\delta}_r=\begin{pmatrix}\delta_r\\0\\\vdots\\0\end{pmatrix},\ r=1,\dots,4\ ,
\end{aligned}
\end{equation}
where $\alpha_1=\alpha_2+\alpha_3+\alpha_4,\ \beta_2=\beta_1+\beta_3+\beta_4,\ \gamma_3=\gamma_1+\gamma_2+\gamma_4, \ \delta_1=\delta_2+\delta_3+\delta_4$. And, we impose following conditions so that the first term belongs to $\mathcal{C}'^{+,\vec{\theta_1}}_{12}$ where $\vec{\theta}_1$ specifies the two dimensional Lorentzian plane in which $P_1$ lies.
\begin{equation}
\begin{aligned}
  \alpha_1<s_1,\ \beta_1>0,\ \gamma_1>0,\ \delta_1<s_1,\ \alpha_1+\beta_1-\delta_1>0,\ \alpha_1+\gamma_1-\delta_1>0\ .
\end{aligned}
\end{equation}
Similarly, in the following we state respectively the conditions that keep the second term inside $\mathcal{C}'^{+,\vec{\theta_2}}_{12}$ ($\vec{\theta}_2$ specifying the two dimensional Lorentzian plane of $P_2$), the third term inside $\mathcal{C}'^{+,\vec{\theta_3}}_{12}$ ($\vec{\theta}_3$ specifying the two dimensional Lorentzian plane of $P_3$), and the fourth term inside $\mathcal{C}'^{+,\vec{\theta_4}}_{12}$ ($\vec{\theta}_2$ specifying the two dimensional Lorentzian plane of $P_{41}$).
\begin{equation}
\begin{aligned}
  &\alpha_2>0,\ \beta_2<s_2,\ \gamma_2>0,\ \delta_2>0,\ \alpha_2+\beta_2-\delta_2<s_2,\ \delta_2-\alpha_2+\gamma_2>0\ ,\\
  &\alpha_3>0,\ \beta_3>0,\ \gamma_3<s_3,\ \delta_3>0,\ \delta_3-\alpha_3+\beta_3>0,\ \alpha_3+\gamma_3-\delta_3<s_3\ ,\\
  &\alpha_4>0,\ \beta_4>0,\ \gamma_4>0,\ \delta_4>0,\ \delta_4-\alpha_4+\beta_4>l_1,\ \delta_4-\alpha_4+\gamma_4>l_1\ .
\end{aligned}
\end{equation}
The above set of conditions along with $s_1,s_2,s_3>0,l_1>0$ can be solved for $\alpha_r,\beta_r,\gamma_r,\delta_r,\ r=1\dots,4$ whenever $s_1,s_2,s_3>l_1$\footnote{Here, the constraints on $s_1,s_2,s_3$ and $l_1$ are derived using Mathematica as an analytical computational tool. We use the same method for the others as well.}. An explicit solution has been given in equation \eqref{eq:QdecompU1expl} in section \ref{sec:case2}.

$\mathcal{B}.2$ \textbf{\textit{Decompositions that involve two $P_{ij}$}}

\noindent In table \ref{tab:decomp2Pij}, we provide the 53 types of decomposition that involve two $P_{ij}$. Their region of validity can be determined using the same procedure as for types of decomposition involving a single $P_{ij}$. In section \ref{sec:numerics} we use only the 13 types which do not involve $P_{14}$ (no. 41 to no. 53 in table \ref{tab:decomp2Pij}).
\begin{table}[H]
\begin{center}
\vspace*{-2.0cm}\hspace*{-1.5cm}\resizebox{20cm}{!}
  {
  \begin{tabular}{||ll|}
    \hline
    & $(P_1,P_2,P_3,P_4)=$\\
    \hline
    1 & $(P_1,0,P_1,P_1)+(0,P_2,0,0)+(0,0,0,-P_1+P_4)+(0,0,-P_1+P_3,0)$\\
\hline
2 & $(P_1,0,0,P_1)+(0,P_2,P_2,0)+(0,0,0,-P_1+P_4)+(0,0,-P_2+P_3,0)$\\
\hline
3 & $(P_1,0,P_1,P_1)+(0,P_2,0,0)+(0,0,-P_1+P_4,-P_1+P_4)+(0,0,P_3-P_4,0)$\\
\hline
4 & $(P_1,P_1,0,P_1)+(0,0,P_3,0)+(0,0,0,-P_1+P_4)+(0,-P_1+P_2,0,0)$\\
\hline
5 & $(P_1,0,0,P_1)+(0,P_3,P_3,0)+(0,0,0,-P_1+P_4)+(0,P_2-P_3,0,0)$\\
\hline
6 & $(P_1,P_1,0,P_1)+(0,0,P_3,0)+(0,-P_1+P_4,0,-P_1+P_4)+(0,P_2-P_4,0,0)$\\
\hline
7 & $(P_1,0,P_1,P_1)+(0,P_2+P_{41},0,0)+(0,P_1-P_4,0,-P_1+P_4)+(0,0,-P_1+P_3,0)$\\
\hline
8 & $(P_1,0,0,P_1)+(0,P_2+P_{41},P_2+P_{41},0)+(0,P_1-P_4,P_1-P_4,-P_1+P_4)+(0,0,-P_2+P_3,0)$\\
\hline
9 & $(P_1,0,P_1,P_1)+(0,P_2+P_{41},0,0)+(0,P_1-P_4,-P_1+P_4,-P_1+P_4)+(0,0,P_3-P_4,0)$\\
\hline
10 & $(P_1,P_1,0,P_1)+(0,0,P_3+P_{41},0)+(0,0,P_1-P_4,-P_1+P_4)+(0,-P_1+P_2,0,0)$\\
\hline
11 & $(P_1,0,0,P_1)+(0,P_3+P_{41},P_3+P_{41},0)+(0,P_1-P_4,P_1-P_4,-P_1+P_4)+(0,P_2-P_3,0,0)$\\
\hline
12 & $(P_1,P_1,0,P_1)+(0,0,P_3+P_{41},0)+(0,-P_1+P_4,P_1-P_4,-P_1+P_4)+(0,P_2-P_4,0,0)$\\
\hline
13 & $(P_2,P_2,0,P_2)+(0,0,P_3,0)+(0,0,0,-P_1+P_4)+(P_1-P_2,0,0,P_1-P_2)$\\
\hline
14 & $(0,P_2,0,0)+(P_3,0,P_3,P_3)+(0,0,0,-P_1+P_4)+(P_1-P_3,0,0,P_1-P_3)$\\
\hline
15 & $(P_2,P_2,0,P_2)+(0,0,P_3,0)+(P_1-P_4,0,0,0)+(-P_2+P_4,0,0,-P_2+P_4)$\\
\hline
16 & $(0,P_2,0,0)+(P_3,0,P_3,P_3)+(P_1-P_4,0,0,0)+(-P_3+P_4,0,0,-P_3+P_4)$\\
\hline
17 & $(0,P_2,0,0)+(P_4,0,P_4,P_4)+(P_1-P_4,0,P_1-P_4,0)+(0,0,-P_1+P_3,0)$\\
\hline
18 & $(0,P_2,P_2,0)+(P_4,0,0,P_4)+(P_1-P_4,0,0,0)+(0,0,-P_2+P_3,0)$\\
\hline
19 & $(0,P_2,0,0)+(P_4,0,P_4,P_4)+(P_1-P_4,0,0,0)+(0,0,P_3-P_4,0)$\\
\hline
20 & $(P_2,P_2,0,P_2)+(0,0,P_3+P_{41},0)+(0,0,P_1-P_4,-P_1+P_4)+(P_1-P_2,0,0,P_1-P_2)$\\
\hline
21 & $(0,P_2,0,0)+(P_3+P_{41},0,P_3+P_{41},P_3+P_{41})+(P_1-P_4,0,P_1-P_4,0)+(P_1-P_3,0,0,P_1-P_3)$\\
\hline
22 & $(P_2,P_2,0,P_2)+(0,0,P_3+P_{41},0)+(P_1-P_4,0,P_1-P_4,0)+(-P_2+P_4,0,0,-P_2+P_4)$\\
\hline
23 & $(0,P_2,0,0)+(P_3+P_{41},0,P_3+P_{41},P_3+P_{41})+(2 (P_1-P_4),0,P_1-P_4,P_1-P_4)+(-P_3+P_4,0,0,-P_3+P_4)$\\
\hline
24 & $(0,0,P_3,0)+(P_4,P_4,0,P_4)+(P_1-P_4,P_1-P_4,0,0)+(0,-P_1+P_2,0,0)$\\
\hline
25 & $(0,P_3,P_3,0)+(P_4,0,0,P_4)+(P_1-P_4,0,0,0)+(0,P_2-P_3,0,0)$\\
\hline
26 & $(0,0,P_3,0)+(P_4,P_4,0,P_4)+(P_1-P_4,0,0,0)+(0,P_2-P_4,0,0)$\\
\hline
27 & $(0,0,P_3,0)+(P_2+P_{41},P_2+P_{41},0,P_2+P_{41})+(P_1-P_4,P_1-P_4,0,0)+(P_1-P_2,0,0,P_1-P_2)$\\
\hline
28 & $(P_3,0,P_3,P_3)+(0,P_2+P_{41},0,0)+(0,P_1-P_4,0,-P_1+P_4)+(P_1-P_3,0,0,P_1-P_3)$\\
\hline
29 & $(0,0,P_3,0)+(P_2+P_{41},P_2+P_{41},0,P_2+P_{41})+(2 (P_1-P_4),P_1-P_4,0,P_1-P_4)+(-P_2+P_4,0,0,-P_2+P_4)$\\
\hline
30 & $(P_3,0,P_3,P_3)+(0,P_2+P_{41},0,0)+(P_1-P_4,P_1-P_4,0,0)+(-P_3+P_4,0,0,-P_3+P_4)$\\
\hline
31 & $(P_4,0,P_4,P_4)+(0,P_2+P_{41},0,0)+(P_1-P_4,P_1-P_4,P_1-P_4,0)+(0,0,-P_1+P_3,0)$\\
\hline
32 & $(P_4,0,0,P_4)+(0,P_2+P_{41},P_2+P_{41},0)+(P_1-P_4,P_1-P_4,P_1-P_4,0)+(0,0,-P_2+P_3,0)$\\
\hline
33 & $(P_4,0,P_4,P_4)+(0,P_2+P_{41},0,0)+(P_1-P_4,P_1-P_4,0,0)+(0,0,P_3-P_4,0)$\\
\hline
34 & $(P_4,P_4,0,P_4)+(0,0,P_3+P_{41},0)+(P_1-P_4,P_1-P_4,P_1-P_4,0)+(0,-P_1+P_2,0,0)$\\
\hline
35 & $(P_4,0,0,P_4)+(0,P_3+P_{41},P_3+P_{41},0)+(P_1-P_4,P_1-P_4,P_1-P_4,0)+(0,P_2-P_3,0,0)$\\
\hline
36 & $(P_4,P_4,0,P_4)+(0,0,P_3+P_{41},0)+(P_1-P_4,0,P_1-P_4,0)+(0,P_2-P_4,0,0)$\\
\hline
37 & $(P_2+P_{41},P_2+P_{41},0,P_2+P_{41})+(0,0,P_3+P_{41},0)+(P_1-P_4,P_1-P_4,P_1-P_4,0)+(P_1-P_2,0,0,P_1-P_2)$\\
\hline
38 & $(0,P_2+P_{41},0,0)+(P_3+P_{41},0,P_3+P_{41},P_3+P_{41})+(P_1-P_4,P_1-P_4,P_1-P_4,0)+(P_1-P_3,0,0,P_1-P_3)$\\
\hline
39 & $(P_2+P_{41},P_2+P_{41},0,P_2+P_{41})+(0,0,P_3+P_{41},0)+(2 (P_1-P_4),P_1-P_4,P_1-P_4,P_1-P_4)+(-P_2+P_4,0,0,-P_2+P_4)$\\
\hline
40 & $(0,P_2+P_{41},0,0)+(P_3+P_{41},0,P_3+P_{41},P_3+P_{41})+(2 (P_1-P_4),P_1-P_4,P_1-P_4,P_1-P_4)+(-P_3+P_4,0,0,-P_3+P_4)$\\
\hline
41 & $(P_1,0,P_1,P_1)+(0,P_2,0,0)+(0,0,-P_1+P_3,-P_1+P_3)+(0,0,0,-P_3+P_4)$\\
\hline
42 & $(P_1,P_1,0,P_1)+(0,0,P_3,0)+(0,-P_1+P_2,0,-P_1+P_2)+(0,0,0,-P_2+P_4)$\\
\hline
43 & $(P_1,P_1,P_1,0)+(0,0,0,P_4)+(0,-P_1+P_2,0,0)+(0,0,-P_1+P_3,0)$\\
\hline
44 & $(P_1,P_1,P_1,0)+(0,0,0,P_4)+(0,-P_1+P_2,-P_1+P_2,0)+(0,0,-P_2+P_3,0)$\\
\hline
45 & $(P_1,P_1,P_1,0)+(0,0,0,P_4)+(0,-P_1+P_3,-P_1+P_3,0)+(0,P_2-P_3,0,0)$\\
\hline
46 & $(P_2,P_2,P_2,0)+(0,0,0,P_2+P_{41})+(P_1-P_2,0,P_1-P_2,P_1-P_2)+(0,0,-P_1+P_3,0)$\\
\hline
47 & $(P_2,P_2,P_2,0)+(0,0,0,P_2+P_{41})+(P_1-P_2,0,0,P_1-P_2)+(0,0,-P_2+P_3,0)$\\
\hline
48 & $(P_2,P_2,P_2,0)+(0,0,0,P_2+P_{41})+(P_1-P_3,0,0,P_1-P_3)+(-P_2+P_3,0,-P_2+P_3,-P_2+P_3)$\\
\hline
49 & $(P_3,P_3,P_3,0)+(0,0,0,P_3+P_{41})+(0,-P_1+P_2,0,0)+(P_1-P_3,P_1-P_3,0,P_1-P_3)$\\
\hline
50 & $(P_3,P_3,P_3,0)+(0,0,0,P_3+P_{41})+(P_1-P_2,0,0,P_1-P_2)+(P_2-P_3,P_2-P_3,0,P_2-P_3)$\\
\hline
51 & $(P_3,P_3,P_3,0)+(0,0,0,P_3+P_{41})+(P_1-P_3,0,0,P_1-P_3)+(0,P_2-P_3,0,0)$\\
\hline
52 & $(P_2+P_{41},P_2+P_{41},0,P_2+P_{41})+(0,0,P_3+P_{41},0)+(2 (P_1-P_2),P_1-P_2,P_1-P_2,P_1-P_2)+(P_2-P_4,P_2-P_4,P_2-P_4,0)$\\
\hline
53 & $(0,P_2+P_{41},0,0)+(P_3+P_{41},0,P_3+P_{41},P_3+P_{41})+(2 (P_1-P_3),P_1-P_3,P_1-P_3,P_1-P_3)+(P_3-P_4,P_3-P_4,P_3-P_4,0)$\\
\hline
  \end{tabular}
  }
\end{center}
\caption{Types of decomposition that involve two $P_{ij}$}
\label{tab:decomp2Pij}
\end{table}
Let us look at one example of the region of validity for the decomposition (no. 44),
\begin{equation}
(P_1,\ P_1,\ P_1,\ 0) + (0,\ 0,\ 0,\ P_4) + (0,\ P_2 - P_1,\ P_2 - P_1,\ 0) + (0,\ 0,\ P_3 - P_2,\ 0)
\end{equation}
from table \ref{tab:decomp2Pij}. The region of validity reads: $P_2^0\geq P_1^0\ \text{and}\ P_3^0\geq P_2^0\ \text{and}$
\begin{eqnarray}
&& (l_4 \leq 0 \cap 
   s_1 > 0 \cap ((0 < s_4 \leq s_1 \cap l_2 < s_4) \cup (s_4 > s_1 \cap 
       l_2 < s_1)))\nonumber\\
&& \cup (l_4 > 0 \cap 
   s_1 > l_4 \cap ((l_4 == s_4 \cap 
       l_2 < 0) \cup (l_2 + l_4 < 
        s_4 \cap (0 < s_4 < l_4 \cup l_4 < s_4 \leq s_1))\nonumber\\
 && \cup (l_2 + l_4 < s_1 \cap 
       s_4 > s_1))) .
\end{eqnarray}
As is clear from the above expression the region of validity includes regions where $l_2$ and/or $l_4$ can be negative. In fact since we are working in a frame where $P_{14}^0=0$ hence this guarantees that $P_{14}$ is space like but there is no restriction on the other $P_{ij}$. This will be true for the cases with three $P_{ij}$ as well. In the following we explicitly give the region of validity for only a few examples from table \ref{tab:decomp2Pij}.\\
\begin{itemize}
\item[41.]{$(P_1,0,P_1,P_1)+(0,P_2,0,0)+(0,0,-P_1+P_3,-P_1+P_3)+(0,0,0,-P_3+P_4)$\\\\
Region of validity: $P_3^0\geq P_1^0$ and $P_4^0\geq P_3^0$ and \\$(l_6 \leq 0 \cap ((s_1 > 0 \cap l_6 + s_1 \leq 0 \cap s_2 > 0 \cap 
       l_3 < s_1) \cup (l_6 + s_1 > 
        0 \cap ((s_2 > 0 \cap l_6 + s_1 \geq 2 s_2 \cap 
           l_3 + l_6 < s_2) \cup (2 s_2 > l_6 + s_1 \cap 
           2 l_3 + l_6 < s_1))))) \cup (l_6 > 0 \cap 
   s_1 > l_6 \cap ((l_6 == s_2 \cap 
       l_3 < 0) \cup (l_3 + l_6 < 
        s_2 \cap (0 < s_2 < 
          l_6 \cup (l_6 < s_2 \cap l_6 + s_1 \geq 2 s_2))) \cup (2 l_3 + l_6 < s_1 \cap 
       2 s_2 > l_6 + s_1)))$.}
\item[43.]{$(P_1,P_1,P_1,0)+(0,0,0,P_4)+(0,-P_1+P_2,0,0)+(0,0,-P_1+P_3,0)$\\\\
Region of validity: $P_2^0\geq P_1^0\ \text{and}\ P_3^0\geq P_1^0$ and $s_1>0 \cap s_4>0 \cap \min\lbrace s_1,s_4\rbrace>\max\lbrace l_2,l_3\rbrace$.}
\item[47.]{$(P_2,P_2,P_2,0)+(0,0,0,P_2+P_{41})+(P_1-P_2,0,0,P_1-P_2)+(0,0,-P_2+P_3,0)$\\\\
Region of validity: $P_1^0\geq P_2^0\ \text{and}\ P_3^0\geq P_2^0$ and $s_2>0 \cap s_5>0 \cap \min\lbrace s_2,s_5\rbrace>\max\lbrace l_2,l_4\rbrace $.}
\end{itemize}

$\mathcal{B}.3$ \textbf{\textit{Decompositions that involve three $P_{ij}$}}

\noindent In table \ref{tab:decomp3Pij}, we provide the 68 types of decomposition that involve three $P_{ij}$. Their region of validity can be determined using the same procedure as for types of decomposition involving a single $P_{ij}$. In section \ref{sec:numerics} we use only the 20 types which do not involve $P_{14}$ (no. 49 to no. 68 in table \ref{tab:decomp3Pij})
\begin{table}[H]
\begin{center}
\vspace*{-2.0cm}\hspace*{-1.5cm}\resizebox{20cm}{!}
  {
  \begin{tabular}{|l|l|}
    \hline
    & $(P_1,P_2,P_3,P_4)=$\\
    \hline
    1 & $(P_1,P_1,P_1,P_1)+(0,0,0,-P_1+P_4)+(0,-P_1+P_2,0,0)+(0,0,-P_1+P_3,0)$\\
\hline
2 & $(P_1,P_1,P_1,P_1)+(0,0,0,-P_1+P_4)+(0,-P_1+P_2,-P_1+P_2,0)+(0,0,-P_2+P_3,0)$\\
\hline
3 & $(P_1,P_1,P_1,P_1)+(0,0,-P_1+P_4,-P_1+P_4)+(0,-P_1+P_2,0,0)+(0,0,P_3-P_4,0)$\\
\hline
4 & $(P_1,P_1,P_1,P_1)+(0,0,0,-P_1+P_4)+(0,-P_1+P_3,-P_1+P_3,0)+(0,P_2-P_3,0,0)$\\
\hline
5 & $(P_1,P_1,P_1,P_1)+(0,-P_1+P_4,0,-P_1+P_4)+(0,0,-P_1+P_3,0)+(0,P_2-P_4,0,0)$\\
\hline
6 & $(P_1,P_1,P_1,P_1)+(0,-P_1+P_4,-P_1+P_4,-P_1+P_4)+(0,0,-P_2+P_3,0)+(0,P_2-P_4,P_2-P_4,0)$\\
\hline
7 & $(P_1,P_1,P_1,P_1)+(0,-P_1+P_4,-P_1+P_4,-P_1+P_4)+(0,P_2-P_3,0,0)+(0,P_3-P_4,P_3-P_4,0)$\\
\hline
8 & $(P_1,P_1,P_1,P_1)+(0,-P_1+P_4,-P_1+P_4,-P_1+P_4)+(0,P_2-P_4,0,0)+(0,0,P_3-P_4,0)$\\
\hline
9 & $(P_2,P_2,P_2,P_2)+(0,0,0,-P_1+P_4)+(P_1-P_2,0,P_1-P_2,P_1-P_2)+(0,0,-P_1+P_3,0)$\\
\hline
10 & $(P_2,P_2,P_2,P_2)+(0,0,0,-P_1+P_4)+(P_1-P_2,0,0,P_1-P_2)+(0,0,-P_2+P_3,0)$\\
\hline
11 & $(P_2,P_2,P_2,P_2)+(0,0,-P_1+P_4,-P_1+P_4)+(P_1-P_2,0,P_1-P_2,P_1-P_2)+(0,0,P_3-P_4,0)$\\
\hline
12 & $(P_2,P_2,P_2,P_2)+(0,0,0,-P_1+P_4)+(P_1-P_3,0,0,P_1-P_3)+(-P_2+P_3,0,-P_2+P_3,-P_2+P_3)$\\
\hline
13 & $(P_2,P_2,P_2,P_2)+(P_1-P_4,0,P_1-P_4,0)+(0,0,-P_1+P_3,0)+(-P_2+P_4,0,-P_2+P_4,-P_2+P_4)$\\
\hline
14 & $(P_2,P_2,P_2,P_2)+(P_1-P_4,0,0,0)+(0,0,-P_2+P_3,0)+(-P_2+P_4,0,0,-P_2+P_4)$\\
\hline
15 & $(P_2,P_2,P_2,P_2)+(P_1-P_4,0,0,0)+(-P_2+P_3,0,-P_2+P_3,-P_2+P_3)+(-P_3+P_4,0,0,-P_3+P_4)$\\
\hline
16 & $(P_2,P_2,P_2,P_2)+(P_1-P_4,0,0,0)+(-P_2+P_4,0,-P_2+P_4,-P_2+P_4)+(0,0,P_3-P_4,0)$\\
\hline
17 & $(P_3,P_3,P_3,P_3)+(0,0,0,-P_1+P_4)+(0,-P_1+P_2,0,0)+(P_1-P_3,P_1-P_3,0,P_1-P_3)$\\
\hline
18 & $(P_3,P_3,P_3,P_3)+(0,0,0,-P_1+P_4)+(P_1-P_2,0,0,P_1-P_2)+(P_2-P_3,P_2-P_3,0,P_2-P_3)$\\
\hline
19 & $(P_3,P_3,P_3,P_3)+(P_1-P_4,P_1-P_4,0,0)+(0,-P_1+P_2,0,0)+(-P_3+P_4,-P_3+P_4,0,-P_3+P_4)$\\
\hline
20 & $(P_3,P_3,P_3,P_3)+(0,0,0,-P_1+P_4)+(P_1-P_3,0,0,P_1-P_3)+(0,P_2-P_3,0,0)$\\
\hline
21 & $(P_3,P_3,P_3,P_3)+(0,-P_1+P_4,0,-P_1+P_4)+(P_1-P_3,P_1-P_3,0,P_1-P_3)+(0,P_2-P_4,0,0)$\\
\hline
22 & $(P_3,P_3,P_3,P_3)+(P_1-P_4,0,0,0)+(P_2-P_3,P_2-P_3,0,P_2-P_3)+(-P_2+P_4,0,0,-P_2+P_4)$\\
\hline
23 & $(P_3,P_3,P_3,P_3)+(P_1-P_4,0,0,0)+(0,P_2-P_3,0,0)+(-P_3+P_4,0,0,-P_3+P_4)$\\
\hline
24 & $(P_3,P_3,P_3,P_3)+(P_1-P_4,0,0,0)+(0,P_2-P_4,0,0)+(-P_3+P_4,-P_3+P_4,0,-P_3+P_4)$\\
\hline
25 & $(P_4,P_4,P_4,P_4)+(P_1-P_4,P_1-P_4,P_1-P_4,0)+(0,-P_1+P_2,0,0)+(0,0,-P_1+P_3,0)$\\
\hline
26 & $(P_4,P_4,P_4,P_4)+(P_1-P_4,P_1-P_4,P_1-P_4,0)+(0,-P_1+P_2,-P_1+P_2,0)+(0,0,-P_2+P_3,0)$\\
\hline
27 & $(P_4,P_4,P_4,P_4)+(P_1-P_4,P_1-P_4,0,0)+(0,-P_1+P_2,0,0)+(0,0,P_3-P_4,0)$\\
\hline
28 & $(P_4,P_4,P_4,P_4)+(P_1-P_4,P_1-P_4,P_1-P_4,0)+(0,-P_1+P_3,-P_1+P_3,0)+(0,P_2-P_3,0,0)$\\
\hline
29 & $(P_4,P_4,P_4,P_4)+(P_1-P_4,0,P_1-P_4,0)+(0,0,-P_1+P_3,0)+(0,P_2-P_4,0,0)$\\
\hline
30 & $(P_4,P_4,P_4,P_4)+(P_1-P_4,0,0,0)+(0,0,-P_2+P_3,0)+(0,P_2-P_4,P_2-P_4,0)$\\
\hline
31 & $(P_4,P_4,P_4,P_4)+(P_1-P_4,0,0,0)+(0,P_2-P_3,0,0)+(0,P_3-P_4,P_3-P_4,0)$\\
\hline
32 & $(P_4,P_4,P_4,P_4)+(P_1-P_4,0,0,0)+(0,P_2-P_4,0,0)+(0,0,P_3-P_4,0)$\\
\hline
33 & $(P_2+P_{41},P_2+P_{41},P_2+P_{41},P_2+P_{41})+(P_1-P_4,P_1-P_4,P_1-P_4,0)+(P_1-P_2,0,P_1-P_2,P_1-P_2)+(0,0,-P_1+P_3,0)$\\
\hline
34 & $(P_2+P_{41},P_2+P_{41},P_2+P_{41},P_2+P_{41})+(P_1-P_4,P_1-P_4,P_1-P_4,0)+(P_1-P_2,0,0,P_1-P_2)+(0,0,-P_2+P_3,0)$\\
\hline
35 & $(P_2+P_{41},P_2+P_{41},P_2+P_{41},P_2+P_{41})+(P_1-P_4,P_1-P_4,0,0)+(P_1-P_2,0,P_1-P_2,P_1-P_2)+(0,0,P_3-P_4,0)$\\
\hline
36 & $(P_2+P_{41},P_2+P_{41},P_2+P_{41},P_2+P_{41})+(P_1-P_4,P_1-P_4,P_1-P_4,0)+(P_1-P_3,0,0,P_1-P_3)+(-P_2+P_3,0,-P_2+P_3,-P_2+P_3)$\\
\hline
37 & $(P_2+P_{41},P_2+P_{41},P_2+P_{41},P_2+P_{41})+(2 (P_1-P_4),P_1-P_4,2 (P_1-P_4),P_1-P_4)+(0,0,-P_1+P_3,0)+(-P_2+P_4,0,-P_2+P_4,-P_2+P_4)$\\
\hline
38 & $(P_2+P_{41},P_2+P_{41},P_2+P_{41},P_2+P_{41})+(2 (P_1-P_4),P_1-P_4,P_1-P_4,P_1-P_4)+(0,0,-P_2+P_3,0)+(-P_2+P_4,0,0,-P_2+P_4)$\\
\hline
39 & $(P_2+P_{41},P_2+P_{41},P_2+P_{41},P_2+P_{41})+(2 (P_1-P_4),P_1-P_4,P_1-P_4,P_1-P_4)+(-P_2+P_3,0,-P_2+P_3,-P_2+P_3)+(-P_3+P_4,0,0,-P_3+P_4)$\\
\hline
40 & $(P_2+P_{41},P_2+P_{41},P_2+P_{41},P_2+P_{41})+(2 (P_1-P_4),P_1-P_4,P_1-P_4,P_1-P_4)+(-P_2+P_4,0,-P_2+P_4,-P_2+P_4)+(0,0,P_3-P_4,0)$\\
\hline
41 & $(P_3+P_{41},P_3+P_{41},P_3+P_{41},P_3+P_{41})+(P_1-P_4,P_1-P_4,P_1-P_4,0)+(0,-P_1+P_2,0,0)+(P_1-P_3,P_1-P_3,0,P_1-P_3)$\\
\hline
42 & $(P_3+P_{41},P_3+P_{41},P_3+P_{41},P_3+P_{41})+(P_1-P_4,P_1-P_4,P_1-P_4,0)+(P_1-P_2,0,0,P_1-P_2)+(P_2-P_3,P_2-P_3,0,P_2-P_3)$\\
\hline
43 & $(P_3+P_{41},P_3+P_{41},P_3+P_{41},P_3+P_{41})+(2 (P_1-P_4),2 (P_1-P_4),P_1-P_4,P_1-P_4)+(0,-P_1+P_2,0,0)+(-P_3+P_4,-P_3+P_4,0,-P_3+P_4)$\\
\hline
44 & $(P_3+P_{41},P_3+P_{41},P_3+P_{41},P_3+P_{41})+(P_1-P_4,P_1-P_4,P_1-P_4,0)+(P_1-P_3,0,0,P_1-P_3)+(0,P_2-P_3,0,0)$\\
\hline
45 & $(P_3+P_{41},P_3+P_{41},P_3+P_{41},P_3+P_{41})+(P_1-P_4,0,P_1-P_4,0)+(P_1-P_3,P_1-P_3,0,P_1-P_3)+(0,P_2-P_4,0,0)$\\
\hline
46 & $(P_3+P_{41},P_3+P_{41},P_3+P_{41},P_3+P_{41})+(2 (P_1-P_4),P_1-P_4,P_1-P_4,P_1-P_4)+(P_2-P_3,P_2-P_3,0,P_2-P_3)+(-P_2+P_4,0,0,-P_2+P_4)$\\
\hline
47 & $(P_3+P_{41},P_3+P_{41},P_3+P_{41},P_3+P_{41})+(2 (P_1-P_4),P_1-P_4,P_1-P_4,P_1-P_4)+(0,P_2-P_3,0,0)+(-P_3+P_4,0,0,-P_3+P_4)$\\
\hline
48 & $(P_3+P_{41},P_3+P_{41},P_3+P_{41},P_3+P_{41})+(2 (P_1-P_4),P_1-P_4,P_1-P_4,P_1-P_4)+(0,P_2-P_4,0,0)+(-P_3+P_4,-P_3+P_4,0,-P_3+P_4)$\\
\hline
49 & $(P_1,P_1,P_1,P_1)+(0,-P_1+P_2,0,-P_1+P_2)+(0,0,-P_1+P_3,0)+(0,0,0,-P_2+P_4)$\\
\hline
50 & $(P_1,P_1,P_1,P_1)+(0,-P_1+P_2,0,0)+(0,0,-P_1+P_3,-P_1+P_3)+(0,0,0,-P_3+P_4)$\\
\hline
51 & $(P_1,P_1,P_1,P_1)+(0,-P_1+P_2,-P_1+P_2,-P_1+P_2)+(0,0,-P_2+P_3,0)+(0,0,0,-P_2+P_4)$\\
\hline
52 & $(P_1,P_1,P_1,P_1)+(0,-P_1+P_2,-P_1+P_2,-P_1+P_2)+(0,0,-P_2+P_3,-P_2+P_3)+(0,0,0,-P_3+P_4)$\\
\hline
53 & $(P_1,P_1,P_1,P_1)+(0,-P_1+P_2,-P_1+P_2,-P_1+P_2)+(0,0,-P_2+P_4,-P_2+P_4)+(0,0,P_3-P_4,0)$\\
\hline
54 & $(P_1,P_1,P_1,P_1)+(0,-P_1+P_3,-P_1+P_3,-P_1+P_3)+(0,P_2-P_3,0,P_2-P_3)+(0,0,0,-P_2+P_4)$\\
\hline
55 & $(P_1,P_1,P_1,P_1)+(0,-P_1+P_3,-P_1+P_3,-P_1+P_3)+(0,P_2-P_3,0,0)+(0,0,0,-P_3+P_4)$\\
\hline
56 & $(P_1,P_1,P_1,P_1)+(0,-P_1+P_3,-P_1+P_3,-P_1+P_3)+(0,P_2-P_4,0,0)+(0,-P_3+P_4,0,-P_3+P_4)$\\
\hline
57 & $(P_2,P_2,P_2,P_2)+(P_1-P_2,0,P_1-P_2,P_1-P_2)+(0,0,-P_1+P_3,-P_1+P_3)+(0,0,0,-P_3+P_4)$\\
\hline
58 & $(P_3,P_3,P_3,P_3)+(0,-P_1+P_2,0,-P_1+P_2)+(P_1-P_3,P_1-P_3,0,P_1-P_3)+(0,0,0,-P_2+P_4)$\\
\hline
59 & $(P_2+P_{41},P_2+P_{41},P_2+P_{41},P_2+P_{41})+(2 (P_1-P_2),P_1-P_2,2 (P_1-P_2),P_1-P_2)+(0,0,-P_1+P_3,0)+(P_2-P_4,P_2-P_4,P_2-P_4,0)$\\
\hline
60 & $(P_2+P_{41},P_2+P_{41},P_2+P_{41},P_2+P_{41})+(2 (P_1-P_2),P_1-P_2,P_1-P_2,P_1-P_2)+(0,0,-P_2+P_3,0)+(P_2-P_4,P_2-P_4,P_2-P_4,0)$\\
\hline
61 & $(P_2+P_{41},P_2+P_{41},P_2+P_{41},P_2+P_{41})+(2 (P_1-P_3),P_1-P_3,P_1-P_3,P_1-P_3)+(-2 (P_2-P_3),-P_2+P_3,-2 (P_2-P_3),-P_2+P_3)+(P_2-P_4,P_2-P_4,P_2-P_4,0)$\\
\hline
62 & $(P_2+P_{41},P_2+P_{41},P_2+P_{41},P_2+P_{41})+(2 (P_1-P_3),P_1-P_3,P_1-P_3,P_1-P_3)+(-P_2+P_3,0,-P_2+P_3,-P_2+P_3)+(P_3-P_4,P_3-P_4,P_3-P_4,0)$\\
\hline
63 & $(P_2+P_{41},P_2+P_{41},P_2+P_{41},P_2+P_{41})+(2 (P_1-P_3),P_1-P_3,P_1-P_3,P_1-P_3)+(-P_2+P_4,0,-P_2+P_4,-P_2+P_4)+(2 (P_3-P_4),P_3-P_4,2 (P_3-P_4),P_3-P_4)$\\
\hline
64 & $(P_3+P_{41},P_3+P_{41},P_3+P_{41},P_3+P_{41})+(0,-P_1+P_2,0,0)+(2 (P_1-P_3),2 (P_1-P_3),P_1-P_3,P_1-P_3)+(P_3-P_4,P_3-P_4,P_3-P_4,0)$\\
\hline
65 & $(P_3+P_{41},P_3+P_{41},P_3+P_{41},P_3+P_{41})+(2 (P_1-P_2),P_1-P_2,P_1-P_2,P_1-P_2)+(P_2-P_3,P_2-P_3,0,P_2-P_3)+(P_2-P_4,P_2-P_4,P_2-P_4,0)$\\
\hline
66 & $(P_3+P_{41},P_3+P_{41},P_3+P_{41},P_3+P_{41})+(2 (P_1-P_2),P_1-P_2,P_1-P_2,P_1-P_2)+(2 (P_2-P_3),2 (P_2-P_3),P_2-P_3,P_2-P_3)+(P_3-P_4,P_3-P_4,P_3-P_4,0)$\\
\hline
67 & $(P_3+P_{41},P_3+P_{41},P_3+P_{41},P_3+P_{41})+(2 (P_1-P_2),P_1-P_2,P_1-P_2,P_1-P_2)+(2 (P_2-P_4),2 (P_2-P_4),P_2-P_4,P_2-P_4)+(-P_3+P_4,-P_3+P_4,0,-P_3+P_4)$\\
\hline
68 & $(P_3+P_{41},P_3+P_{41},P_3+P_{41},P_3+P_{41})+(2 (P_1-P_3),P_1-P_3,P_1-P_3,P_1-P_3)+(0,P_2-P_3,0,0)+(P_3-P_4,P_3-P_4,P_3-P_4,0)$\\
\hline
  \end{tabular}
  }
\end{center}
\caption{Types of decomposition that involve three $P_{ij}$}
\label{tab:decomp3Pij}
\end{table}
Here also we provide only a few examples from table \ref{tab:decomp3Pij}.\\
\begin{itemize}
\item[1.]{$(P_1,P_1,P_1,P_1)+(0,0,0,-P_1+P_4)+(0,-P_1+P_2,0,0)+(0,0,-P_1+P_3,0)$\\\\
Region of validity: $P_2^0\geq P_1^0$ and $P_3^0\geq P_1^0$ and $s_1>l_1>0 \cap s_1>\max\lbrace l_1+l_2,\ l_1+l_3\rbrace$.}
\item[49.]{$(P_1,P_1,P_1,P_1)+(0,-P_1+P_2,0,-P_1+P_2)+(0,0,-P_1+P_3,0)+(0,0,0,-P_2+P_4)$\\\\ 
Region of validity: $P_2^0\geq P_1^0$ and $P_3^0\geq P_1^0$ and $P_4^0\geq P_2^0$ and\\
$(l_3 < s_1 \cap ((l_5 < 
        0 \cap ((s_1 > 0 \cap l_2 < s_1 \cap l_5 + s_1 \leq 0) \cup (l_5 + s_1 > 0 \cap 
           3 l_5 + 2 s_1 < 0 \cap l_2 + l_5 \leq 0) \cup (3 l_5 + 2 s_1 \geq 0 \cap 
           l_2 + l_5 < 0))) \cup (s_1 > l_5 \cap l_5 \geq 0 \cap 
       l_2 + l_5 < 0))) \cup (2 l_2 + l_5 < s_1 \cap 
   l_2 + l_3 + l_5 < 
    s_1 \cap ((l_5 < 
        0 \cap ((l_2 + l_5 > 0 \cap l_5 + s_1 > 0 \cap 
           3 l_5 + 2 s_1 < 0) \cup (l_2 + l_5 \geq 0 \cap 
           3 l_5 + 2 s_1 \geq 0))) \cup (s_1 > l_5 \cap l_5 \geq 0 \cap 
       l_2 + l_5 \geq 0)))$.}
\item[50.]{$(P_1,P_1,P_1,P_1)+(0,-P_1+P_2,0,0)+(0,0,-P_1+P_3,-P_1+P_3)+(0,0,0,-P_3+P_4)$\\\\
Region of validity: $P_2^0\geq P_1^0$ and $P_3^0\geq P_1^0$ and $P_4^0\geq P_3^0$ and\\
$(l_2 < s_1 \cap ((l_6 < 
        0 \cap ((s_1 > 0 \cap l_3 < s_1 \cap l_6 + s_1 \leq 0) \cup (l_6 + s_1 > 0 \cap 
           3 l_6 + 2 s_1 < 0 \cap l_3 + l_6 \leq 0) \cup (3 l_6 + 2 s_1 \geq 0 \cap 
           l_3 + l_6 < 0))) \cup (s_1 > l_6 \cap l_6 \geq 0 \cap 
       l_3 + l_6 < 0))) \cup (2 l_3 + l_6 < s_1 \cap 
   l_2 + l_3 + l_6 < 
    s_1 \cap ((l_6 < 
        0 \cap ((l_3 + l_6 > 0 \cap l_6 + s_1 > 0 \cap 
           3 l_6 + 2 s_1 < 0) \cup (l_3 + l_6 \geq 0 \cap 
           3 l_6 + 2 s_1 \geq 0))) \cup (s_1 > l_6 \cap l_6 \geq 0 \cap 
       l_3 + l_6 \geq 0)))$.}
\end{itemize}

\section{Improving the algorithm}
\label{app:algo2}

The general strategy to improve our algorithm is to work with a list other than \eqref{eq:algostep2} which is,
\begin{equation}
\label{eq:algo2.1}
  \bigg\{P_i,\ P_2+P_4-P_1,\ P_3+P_4-P_1,\ P_{ij},\ i\neq j,\ i,j=1,\dots,4\bigg\}\ .
\end{equation}
Notice that this is not the only possibility. We could extend the list by including all combinations obtained by taking the sum of two $P_{ij}$ at a time and avoiding any repetitions. By this we mean that we can keep $P_{12}+P_{13}=2P_1-P_2-P_3$ but not $P_{12}+P_{24}=P_1-P_4=P_{14}$ since the latter is already included. Instead of stopping at two one could also consider sums of three $P_{ij}$. So there are many such possibilities and hence many more decompositions can be obtained each having some region of validity.

For example consider the following two new decompositions apart from the ones obtained from our algorithm.
\begin{equation}
(P_1,P_2,P_3,P_4)=(P_1, 0, 0, P_1) + (0, P_2, P_2, 0) + (0, 0, P_{32}, P_{32}) + (0, 0, 0,P_{21}+ P_{43}),
\end{equation}
with region of validity: $P_3^0\geq P_2^0$ and $P_2^0+P_4^0\geq P_1^0+P_3^0$ and\\
$(l_7 \leq 0 \cap ((s_1 > 0 \cap 
       l_7 + s_1 \leq 
        0 \cap ((s_2 > 0 \cap 2 l_4 < s_1 + s_2 \cap s_2 \leq s1) \cup (s_2 > s_1 \cap 
           l_4 < s_1))) \cup (l_7 + s_1 > 
        0 \cap ((0 < s2 \leq l_7 + 2 s_1 \cap 
           2 l_4 + l_7 < s_2) \cup (s_2 > l_7 + 2 s_1 \cap l_4 < s1))))) \cup (l_7 >
     0 \cap 
   s_1 > l_7 \cap ((0 < s_2 < l_7 \cap l_4 + l_7 < s_2) \cup (l_7 == s_2 \cap 
       l_4 < 0) \cup (l_7 < s_2 \cap l_7 + s_2 \leq 2 s_1 \cap 
       2 l_4 + l_7 < s_2) \cup (l_7 + s_2 > 2 s_1 \cap l_4 + l_7 < s_1)))$.
\begin{equation}
(P_1,P_2,P_3,P_4)=(P_1, 0, 0, P_1) + (0, P_3, P_3, 0) + (0, P_{23}, 0, P_{23}) + (0, 0, 0,P_{31}+ P_{42}),
\end{equation}
with region of validity: $P_2^0\geq P_3^0$ and $P_3^0+P_4^0\geq P_1^0+P_2^0$ and\\
$(l_4 \leq 0 \cap ((s_1 > 0 \cap l_4 + s_1 \leq 0 \cap s_3 > 0 \cap 
       l_8 < s_1) \cup (l_4 + s_1 > 
        0 \cap ((0 < s_3 \leq l_4 + s_1 \cap l_4 + l_8 < s_3) \cup (s_3 > l_4 + s_1 \cap 
           l_8 < s_1))))) \cup (l_4 > 0 \cap 
   s_1 > l_4 \cap ((2 l_4 == s_3 \cap 
       l_8 < 0) \cup (2 l_4 + l_8 < 
        s_3 \cap (l_4 < s_3 < 2 l_4 \cup 2 l_4 < s_3 \leq l_4 + s_1)) \cup (l_4 + l_8 < 
        s_1 \cap s_3 > l_4 + s_1)))$. \\
Here we have defined,
\begin{equation}
l_7=\|\vec{P}_{21}+\vec{P}_{43}\|-|P_{21}^0+P_{43}^0|\ \ \text{and}\ \ l_8=\|\vec{P}_{31}+\vec{P}_{42}\|-|P_{31}^0+P_{42}^0|\ .
\end{equation}
For the case of $P_2^0=P_3^0=1$ this inclusion changes the percentage of points covered from $15.8916\%$ to $16.552\%$. 
It is because of this reason we naively hope that the convex hull of these remaining 20 LES tubes may yield the corresponding primitive tube completely.

\end{document}